\documentclass[letterpaper, journal, 10pt]{IEEEtran}

\usepackage{etex}
\usepackage{cite}
\usepackage{nicefrac}
\usepackage{pgf,tikz}
\usepackage{pgfplots}
\usetikzlibrary{calc,shapes,positioning,arrows,decorations.markings,fit,plotmarks}
\usepackage[usestackEOL]{stackengine} 

\usepackage[cmex10]{amsmath}
\usepackage{amssymb}
\usepackage[T1]{fontenc} 

\usepackage{bm}
\usepackage[varg]{txfonts}
\let\mathbb=\varmathbb
\DeclareSymbolFont{letters}{OML}{ztmcm}{m}{it}
\usepackage[font=footnotesize]{subfig}

\usepackage{tabularx,colortbl}
\usepackage{multirow}
\usepackage{dcolumn}
\usepackage{booktabs}
\newcolumntype{C}[1]{>{\centering\arraybackslash}p{#1}}

\usepackage{color}
\usepackage{hyphenat}
\usepackage{xspace}
\newcommand{\polarbear}{\nohyphens{\textsc{PolarBear}}\xspace}

\makeatletter

\pgfdeclareshape{reg}{
  \savedanchor\northeast{%
    \pgfmathsetlength\pgf@x{\pgfshapeminwidth}%
    \pgfmathsetlength\pgf@y{\pgfshapeminheight}%
    \pgf@x=0.325\pgf@x
    \pgf@y=0.325\pgf@y
  }
  \savedanchor\southwest{%
    \pgfmathsetlength\pgf@x{\pgfshapeminwidth}%
    \pgfmathsetlength\pgf@y{\pgfshapeminheight}%
    \pgf@x=-0.325\pgf@x
    \pgf@y=-0.325\pgf@y
  }
  \inheritanchorborder[from=rectangle]

  \anchor{center}{\pgfpointorigin}
  \anchor{north}{\northeast \pgf@x=0pt}
  \anchor{east}{\northeast \pgf@y=0pt}
  \anchor{south}{\southwest \pgf@x=0pt}
  \anchor{west}{\southwest \pgf@y=0pt}
  \anchor{north east}{\northeast}
  \anchor{north west}{\northeast \pgf@x=-\pgf@x}
  \anchor{south west}{\southwest}
  \anchor{south east}{\southwest \pgf@x=-\pgf@x}
  \anchor{text}{
    \pgfpointorigin
    \advance\pgf@x by -.5\wd\pgfnodeparttextbox%
    \advance\pgf@y by -.5\ht\pgfnodeparttextbox%
    \advance\pgf@y by +.5\dp\pgfnodeparttextbox%
  }

  \anchor{D}{
    \pgf@process{\northeast}%
    \pgf@x=-1\pgf@x%
    \pgf@y=.5\pgf@y%
  }
  \anchor{CLK}{
    \pgf@process{\northeast}%
    \pgf@x=-1\pgf@x%
    \pgf@y=-.66666\pgf@y%
  }
  \anchor{Q}{
    \pgf@process{\northeast}%
    \pgf@y=.5\pgf@y%
  }
  \backgroundpath{
    \pgfpathrectanglecorners{\southwest}{\northeast}
    \pgf@anchor@reg@CLK
    \pgf@xa=\pgf@x \pgf@ya=\pgf@y
    \pgf@xb=\pgf@x \pgf@yb=\pgf@y
    \pgf@xc=\pgf@x \pgf@yc=\pgf@y
    \pgfmathsetlength\pgf@x{0.5ex} 
    \advance\pgf@ya by \pgf@x
    \advance\pgf@xb by \pgf@x
    \advance\pgf@yc by -\pgf@x
    \pgfpathmoveto{\pgfpoint{\pgf@xa}{\pgf@ya}}
    \pgfpathlineto{\pgfpoint{\pgf@xb}{\pgf@yb}}
    \pgfpathlineto{\pgfpoint{\pgf@xc}{\pgf@yc}}
    \pgfclosepath

  }
}

\tikzset{add font/.code={\expandafter\def\expandafter\tikz@textfont\expandafter{\tikz@textfont#1}}} 

\tikzset{flip flop/port labels/.style={font=\sffamily\scriptsize}}
\tikzset{every reg node/.style={draw,fill=blue!10,minimum width=0.75cm,minimum height=1.2cm,thick,inner sep=1mm,outer sep=0pt,cap=round,add font=\sffamily\scriptsize}}

\makeatother

\newcommand{\ub}[1]{$#1$}
\newcommand{\fb}[1]{\color{gray}$#1$}
\definecolor{darkgreen}{RGB}{0, 128, 0}
\definecolor{mypurple}{RGB}{153, 71, 155}

\usepackage{glossaries}

\usepackage{ifpdf}
\ifpdf
\usepackage[draft,pdfborder={0 0 0}]{hyperref}
\fi

\begin{document}
\bstctlcite{IEEEexample:BSTcontrol}

\title{\polarbear: A 28-nm FD-SOI ASIC\\for Decoding of Polar Codes}

\author{Pascal Giard,~\IEEEmembership{Member,~IEEE}, Alexios Balatsoukas-Stimming, Thomas Christoph M{\"u}ller,~\IEEEmembership{Student~Member,~IEEE},\\%
  Andrea Bonetti,~\IEEEmembership{Student~Member,~IEEE}, Claude Thibeault,~\IEEEmembership{Senior~Member,~IEEE},\\%
  Warren J. Gross,~\IEEEmembership{Senior~Member,~IEEE}, Philippe Flatresse, and Andreas Burg,~\IEEEmembership{Member,~IEEE}
\thanks{P. Giard, A. Balatsoukas-Stimming, T.~C. M\"uller, A. Bonetti, and A. Burg are with the Telecommunications Circuits Laboratory, \'Ecole polytechnique f\'ed\'erale de Lausanne, 1015 Lausanne, VD, Switzerland (e-mail: \{pascal.giard, alexios.balatsoukas,christoph.mueller,andrea.bonetti,andreas.burg\}@epfl.ch).}%
\thanks{C. Thibeault is with the Department of Electrical Engineering, \'Ecole de technologie sup\'erieure, Montr\'eal, QC, H3C 1K3, Canada (e-mail: claude.thibeault@etsmtl.ca).}%
\thanks{W. J. Gross is with the Department of Electrical and Computer Engineering, McGill University, Montr\'eal, QC, H3A 0G4, Canada (e-mail: warren.gross@mcgill.ca).}%
\thanks{P. Flatresse is with STMicroelectronics, 38920 Crolles, France.}%
}

\newacronym[plural=CRCs]{crc}{CRC}{cyclic redundancy check}
\newacronym{fer}{FER}{frame-error rate}
\newacronym{ber}{BER}{bit-error rate}
\newacronym{ldpc}{LDPC}{low-density parity-check}
\newacronym{sc}{SC}{successive-cancellation}
\newacronym{bp}{BP}{belief-propagation}
\newacronym{bpsk}{BPSK}{binary phase-shift keying}
\newacronym{awgn}{AWGN}{additive white Gaussian noise}
\newacronym[plural=LLRs,firstplural=log-likelihood ratios (LLRs)]{llr}{LLR}{log-likelihood-ratio}
\newacronym{scl}{SCL}{successive-cancellation list}
\newacronym{scf}{SCF}{successive-cancellation flip}
\newacronym{spc}{SPC}{single-parity check}
\newacronym{fll}{FLL}{frequency lock loop}
\newacronym{cgu}{CGU}{clock-generation unit}
\newacronym{tcu}{TCU}{test-controller unit}
\newacronym[plural=CCs,firstplural=clock cycles (CCs)]{cc}{CC}{clock cycle}
\newacronym[plural=PSNs,firstplural=partial-sum networks (PSNs)]{psn}{PSN}{partial-sum network}
\newacronym{fsm}{FSM}{finite-state machine}
\newacronym{wc}{W.-C.}{worst-case}

\maketitle

\begin{abstract}
Polar codes are a recently proposed class of block codes that provably achieve the capacity of various communication channels. They received a lot of attention as they can do so with low-complexity encoding and decoding algorithms, and they have an explicit construction. Their recent inclusion in a 5G communication standard will only spur more research. However, only a couple of ASICs featuring decoders for polar codes were fabricated, and none of them implements a list-based decoding algorithm. In this paper, we present ASIC measurement results for a fabricated 28\,nm CMOS chip that implements two different decoders: the first decoder is tailored toward error-correction performance and flexibility. It supports any code rate as well as three different decoding algorithms: successive cancellation (SC), SC flip and SC list (SCL). The flexible decoder can also decode both non-systematic and systematic polar codes. The second decoder targets speed and energy efficiency. We present measurement results for the first silicon-proven SCL decoder, where its coded throughput is shown to be of 306.8\,Mbps with a latency of 3.34\,us and an energy per bit of 418.3\,pJ/bit at a clock frequency of 721\,MHz for a supply of 1.3\,V. The energy per bit drops down to 178.1\,pJ/bit with a more modest clock frequency of 308\,MHz, lower throughput of 130.9\,Mbps and a reduced supply voltage of 0.9\,V. For the other two operating modes, the energy per bit is shown to be of approximately 95\,pJ/bit. The less flexible high-throughput unrolled decoder can achieve a coded throughput of 9.2\,Gbps and a latency of 628\,ns for a measured energy per bit of 1.15\,pJ/bit at 451\,MHz.
\end{abstract}


\section{Introduction}
\IEEEPARstart{P}{olar} codes~\cite{Arikan2009} received a lot of attention in the recent years, and they will gather even more as they have just been selected for the 5G communication standard currently under development by the 3GPP~\cite[p.\,139]{3GPPRANPolar}. However, to this day, only a couple of ASICs featuring decoders for polar codes have been fabricated~\cite{Mishra2012,Park2014}, making it difficult to get a good picture of what can be achieved. The chip described in \cite{Mishra2012} is for a \gls{sc} decoder that lacks the very significant algorithmic and error-correction performance improvements that were later added to the basic \gls{sc} algorithm, e.g., \cite{Alamdar-Yazdi2011,Sarkis_JSAC_2014,Tal2015}, and was fabricated on outdated technology node which does not suffer from the physical post-layout limitations of modern processes. The chip presented in \cite{Park2014} was built for a more recent technology but solely implements the belief-propagation decoding, an algorithm that, even compared to \gls{sc}, suffers from mediocre error-correction performance at short to moderate blocklength.

Moreover, \gls{scl} is regarded as the most promising decoding algorithm, yet, up to now it has not been silicon proven. \Gls{scf} decoding is another promising algorithm~\cite{Afisiadis2014} for applications that can tolerate a variable decoding throughput for the benefit of superior energy efficiency. However, it has never been implemented in hardware before.

\subsubsection*{Contributions} In this paper, we present and compare two very different architectural choices for decoding of polar codes: flexible and optimized for error-correction performance versus high speed and good energy efficiency. We introduce a simple latency saving technique that is directly applicable to the \gls{sc}, \gls{scf}, and \gls{scl} decoding algorithms. We describe a flexible decoder that supports any code rate for any set of frozen-bit locations as well as three different decoding algorithms with parameters that are configurable at the time of execution. Furthermore, this flexible decoder can decode both non-systematic and systematic polar codes. We present the first hardware implementation of the \gls{scf} algorithm along with its corresponding measurement results, and we show with measurement results that a dedicated fully-unrolled \gls{sc} decoder offers the best energy efficiency that is almost two orders of magnitude better than a sequential list decoder. This points out the substantial cost for improving error-correction performance beyond \gls{sc} decoding and for providing flexibility.

\subsubsection*{Outline}
The remainder of this paper starts with Section~\ref{sec:background} which provides the necessary background about polar codes along with a brief overview of the various decoding algorithms implemented on our fabricated chip. The impact on the error-correction performance of these different algorithms is also illustrated in that section. Section~\ref{sec:chip_architecture} describes the architecture of the \polarbear chip, including the hardware implementations of the two decoders with entirely orthogonal objectives featured on the chip, and the units that are necessary for the chip to function properly and to be testable. Section~\ref{sec:results} shows how the various modes of the flexible decoder compare and presents the advantages and disadvantages of each, and similarly for the two architectural directions. For that purpose, detailed measurement results are presented and discussed for each decoder. A comparison against the state-of-the-art fabricated polar decoders is also carried out in that section. Finally, Section~\ref{sec:conclusion} concludes this paper.

\vspace{-3pt}
\section{Polar Codes}\label{sec:background}
\subsection{Construction and Encoding}
In his seminal work on polar codes \cite{Arikan2009}, Ar{\i}kan showed that using a particular linear transformation on a vector of bits leads to a polarization phenomenon, where some of the bits become almost completely reliable when transmitted over certain types of channels while the remainder become almost completely unreliable. Polar codes exploit this phenomenon, thus provably achieving the symmetric capacity of memoryless channels as the blocklength grows to infinity.

An ($N$, $k$) polar code has a blocklength of $N$ and rate $R=\frac{k}{N}$. It is constructed by setting the $N-k$ least reliable bits---called frozen bits---of a row vector $\bm{u}$ of length $N$ to a predetermined value, typically zero, while the remaining $k$ locations in $\bm{u}$ are used to carry the information bits $a_i$, $0 \leq i < k$. The set of frozen-bit indices is denoted by $\mathcal{A}^{\text{c}}$ and the set of information indices is denoted by $\mathcal{A}$. The encoding process consists in multiplying this row vector $\bm{u}$ by a $N \times N$ generator matrix $\bm{F}^{\otimes n}$, where $\bm{F}^{\otimes n}$ is recursively defined as:
\begin{equation}
	\bm{F}^{\otimes n} = \left[ \begin{smallmatrix} \bm{F}^{\otimes(n-1)} & 0 \\ \bm{F}^{\otimes(n-1)} & \bm{F}^{\otimes(n-1)}\end{smallmatrix} \right]
\end{equation}
with $\otimes n$ denoting the $n$-th Kronecker product of the Ar{\i}kan kernel matrix $\bm{F}^{\otimes 1}=\bm{F}=\left[ \begin{smallmatrix} 1 & 0 \\ 1 & 1\end{smallmatrix} \right]$, and $n=\log_2(N)$.

\begin{figure}[t]
  \centering
  \begin{tikzpicture}
    \definecolor{var_fill}{RGB}{0,0,0}
    \definecolor{chk_fill}{RGB}{255,255,255}

    \tikzset{
      chk/.style={draw,fill=chk_fill,circle,minimum size=4mm, inner sep=0},
      var/.style={draw,fill=var_fill,circle,minimum size=1mm, inner sep=0},
      sep/.style={rectangle,minimum width=4mm, inner sep=0},
      empty/.style={rectangle, inner sep=0},
      bit/.style={rectangle, inner sep = 1mm}
    }

    \matrix[row sep=2mm, column sep=2mm] {
      \node[bit] (n0s0) {\fb{u_0=0}}; & \node[sep] {}; & \node[chk] (n0s1) {$+$}; & \node[sep] (s10) {}; & \node[chk] (n0s2) {$+$}; & \node[empty] {};              & \node[sep] (s20) {}; & \node[chk] (n0s3) {$+$}; & \node[empty] {}; & \node[empty] {}; & \node[empty] {}; && \node[bit] (xn0s4) {\ub{x_0}};\\
      \node[bit] (n1s0) {\fb{u_1=0}}; & \node[sep] {}; & \node[var] (n1s1) {};    & \node[sep] (s11) {}; &                              & \node[chk] (n1s2) {$+$};  & \node[sep] (s21) {}; & \node[empty] {};             & \node[chk] (n1s3) {$+$}; & \node[empty] {}; & \node[empty] {}; && \node[bit] (xn1s4) {\ub{x_1}};\\
      \node[bit] (n2s0) {\fb{u_2=0}}; & \node[sep] {}; & \node[chk] (n2s1) {$+$}; & \node[sep] (s12) {}; & \node[var] (n2s2) {};    & \node[empty] {};              & \node[sep] (s22) {}; & \node[empty] {};             & \node[empty] {}; & \node[chk] (n2s3) {$+$}; & \node[empty] {}; && \node[bit] (xn2s4) {\ub{x_2}};\\

      \node[bit] (n3s0) {\ub{u_3=a_0}}; & \node[sep] {}; & \node[var] (n3s1) {};    & \node[sep] (s13) {}; & \node[empty] {};             & \node[var] (n3s2) {};     & \node[sep] (s23) {}; & \node[empty] {};             & \node[empty] {}; & \node[empty] {}; & \node[chk] (n3s3) {$+$}; && \node[bit] (xn3s4) {\ub{x_3}};\\

      \node[bit] (n4s0) {\fb{u_4=0}}; & \node[sep] {}; & \node[chk] (n4s1) {$+$}; & \node[sep] (s14) {}; & \node[chk] (n4s2) {$+$}; & \node[empty] {};              & \node[sep] (s24) {}; & \node[var] (n4s3) {};    & \node[empty] {}; & \node[empty] {}; & \node[empty] {}; && \node[bit] (xn4s4) {\ub{x_4}};\\
      \node[bit] (n5s0) {\ub{u_5=a_1}}; & \node[sep] {}; & \node[var] (n5s1) {};    & \node[sep] (s15) {}; &                              & \node[chk] (n5s2) {$+$};  & \node[sep] (s25) {}; & \node[empty] {};             & \node[var] (n5s3) {}; & \node[empty] {}; &  \node[empty] {}; && \node[bit] (xn5s4) {\ub{x_5}};\\
      \node[bit] (n6s0) {\ub{u_6=a_2}}; & \node[sep] {}; & \node[chk] (n6s1) {$+$}; & \node[sep] (s16) {}; & \node[var] (n6s2) {};    & \node[empty] {};              & \node[sep] (s26) {}; & \node[empty] {};             & \node[empty] {}; & \node[var] (n6s3) {}; &  \node[empty] {}; && \node[bit] (xn6s4) {\ub{x_6}};\\
      
      \node[bit] (n7s0) {\ub{u_7=a_3}}; & \node[sep] {}; & \node[var] (n7s1) {};    & \node[sep] (s17) {}; &                              & \node[var] (n7s2) {};  & \node[sep] (s27) {}; & \node[empty] {};             & \node[empty] {}; & \node[empty] {}; &  \node[var] (n7s3) {}; && \node[bit] (xn7s4) {\ub{x_7}};\\
    };
    \path[-] (n0s0) edge (n0s1) (n0s1) edge (n0s2) (n0s2) edge (n0s3) (n0s3) edge (xn0s4);
    \path[-] (n1s0) edge (n1s1) (n1s1) edge (n1s2) (n1s2) edge (n1s3) (n1s3) edge (xn1s4);
    \path[-] (n2s0) edge (n2s1) (n2s1) edge (n2s2) (n2s2) edge (n2s3) (n2s3) edge (xn2s4);
    \path[-] (n3s0) edge (n3s1) (n3s1) edge (n3s2) (n3s2) edge (n3s3) (n3s3) edge (xn3s4);
    \path[-] (n4s0) edge (n4s1) (n4s1) edge (n4s2) (n4s2) edge (n4s3) (n4s3) edge (xn4s4);
    \path[-] (n5s0) edge (n5s1) (n5s1) edge (n5s2) (n5s2) edge (n5s3) (n5s3) edge (xn5s4);
    \path[-] (n6s0) edge (n6s1) (n6s1) edge (n6s2) (n6s2) edge (n6s3) (n6s3) edge (xn6s4);
    \path[-] (n7s0) edge (n7s1) (n7s1) edge (n7s2) (n7s2) edge (n7s3) (n7s3) edge (xn7s4);

    \path[-] (n0s1) edge (n1s1);
    \path[-] (n2s1) edge (n3s1);
    \path[-] (n4s1) edge (n5s1);
    \path[-] (n6s1) edge (n7s1);

    \path[-] (n0s2) edge (n2s2);
    \path[-] (n1s2) edge (n3s2);
    \path[-] (n4s2) edge (n6s2);
    \path[-] (n5s2) edge (n7s2);

    \path[-] (n0s3) edge (n4s3);
    \path[-] (n1s3) edge (n5s3);
    \path[-] (n2s3) edge (n6s3);
    \path[-] (n3s3) edge (n7s3);

  \end{tikzpicture}
  \caption{Graph representation of a (8, 4) polar code.}
  \label{fig:pc8-enc}
\end{figure}
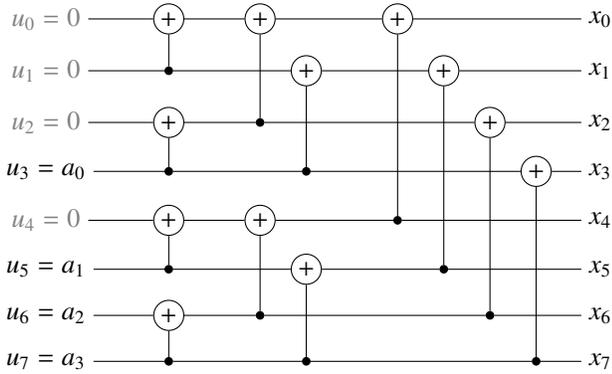

Fig.~\ref{fig:pc8-enc} illustrates the encoding process as a graph where $\oplus$ represents a modulo-2 addition (XOR). In that representation, a polar codeword is generated by setting the frozen- and information-bit locations to 0 and $a_i$, $0 \leq i < k$, respectively, on the left and by propagating data through the graph from left to right.
Polar codes can also be encoded systematically as described and efficiently implemented in \cite{Arikan2011} and \cite{Sarkis2016}, respectively.
Systematic and non-systematic polar codes have the same \gls{fer}. In this paper, unless otherwise specified, non-systematic polar coding is used.

\subsection{Successive-Cancellation (SC) Decoding}
The \gls{sc} decoding algorithm as initially proposed~\cite{Arikan2009} proceeds by visiting the graph representation of Fig.~\ref{fig:pc8-enc} sequentially, from right to left, from top to bottom, successively estimating $\bm{\hat{u}}$ from the noisy channel values. To reduce latency and increase throughput, it was first proposed to calculate two bits at once \cite{Mishra2012}. Later, the \gls{sc} algorithm was further refined to use the a priori knowledge of the frozen bit locations to trim the graph \cite{Alamdar-Yazdi2011} or even to use dedicated, and faster, decoding algorithms on parts of the graph \cite{Sarkis_JSAC_2014}. Regardless of the version of the \gls{sc} algorithm used, at all times, only one candidate codeword is considered.

\subsection{Successive-Cancellation Flip (SCF) Decoding}\label{sec:bg:scf}
The \gls{scf} decoding algorithm \cite{Afisiadis2014} shares many similarities with the \gls{sc} algorithm. Initially, it proceeds exactly like \gls{sc} decoding but while decoding it also keeps a list of the least reliable bit-decisions. Moreover it is necessary to concatenate a \gls{crc} with the polar code. Once the \gls{scf} decoder has generated a complete codeword candidate, it checks if the calculated \gls{crc} matches the expected one. If the \gls{crc} check fails, then \gls{sc} decoding is restarted until the bit corresponding to the least reliable bit-decision is reached. Once reached, the \gls{scf} flips that decision and resumes \gls{sc} decoding. After this second round, if the calculated \gls{crc} still does not match the expected \gls{crc}, then the algorithm is rerun once more and the second least reliable bit-decision is flipped. This procedure lasts until the \gls{crc} comparison succeeds or until the maximum number of trials is reached.

\subsection{Successive-Cancellation List (SCL) Decoding}\label{sec:bg:scl}
As the name indicates, the \gls{scl} algorithm \cite{Tal2015} also shares many similarities with the \gls{sc} algorithm. Contrary to \gls{sc} decoding though, the \gls{scl} decoding algorithm builds a constrained list of up to $L$ of candidate codewords. It does so by examining both possibilities of $\hat{u}_i$ for the locations $i$ corresponding to information bits. A path reliability metric, calculated along the way, is used to keep only the $L$-best paths in the survivor list. At the very end of the decoding process, the candidate with the best path reliability metric among the $L$ candidates is picked as the estimated codeword.

If a polar code is concatenated with a \gls{crc}, the \gls{crc} for each of the $L$ candidates is calculated and compared against the expected one. The most reliable candidate out of all candidates that pass the CRC is selected as the decoded codeword. If all candidates fail the CRC, then the algorithm simply picks the candidate with the best path reliability metric. In this work, all \gls{scl} results use an 8-bit \gls{crc}.

\subsection{Error-Correction Performance Comparison}

\begin{figure}[t]
	\centering
  \begin{tikzpicture}

    \pgfplotsset{
      grid style = {
        dash pattern = on 0.05mm off 1mm,
        line cap = round,
        black,
        line width = 0.5pt
      },
      label style = {font=\fontsize{9pt}{7.2}\selectfont},
      tick label style = {font=\fontsize{9pt}{7.2}\selectfont}
    }

    \begin{semilogyaxis}[%
      xlabel=$E_b/N_0$ (dB),xtick={2,3,...,6.0},%
      xlabel style={yshift=0.4em},%
      minor x tick num={1},
      xmin=2.5,xmax=6,%
      ymin=1e-5,ymax=1e0,%
      ylabel=Frame-error rate, ylabel style={yshift=-0.6em},%
      width=0.9\columnwidth, height=6.15cm, grid=major,%
      legend style={
        anchor={center},
        cells={anchor=west},
        column sep=1.5mm,
        font=\fontsize{9pt}{7.2}\selectfont,
        mark size=3.0pt,
        mark options=solid
      },
      legend columns=5,
      legend to name=ec-perf-legend,
      mark size=3.0pt,
      mark options=solid]
      
      \addlegendimage{empty legend}
      \addlegendentry[anchor=east]{SC\phantom{F}:}

      \addplot[very thick,color=black] plot coordinates {
        (2.00000e+00, 9.98879e-01)
        (2.25000e+00, 9.90935e-01)
        (2.50000e+00, 9.56785e-01)
        (2.75000e+00, 8.63794e-01)
        (3.00000e+00, 6.88224e-01)
        (3.25000e+00, 4.67271e-01)
        (3.50000e+00, 2.59869e-01)
        (3.75000e+00, 1.21738e-01)
        (4.00000e+00, 4.80000e-02)
        (4.25000e+00, 1.70467e-02)
        (4.50000e+00, 6.20561e-03)
        (4.75000e+00, 1.80531e-03)
        (5.00000e+00, 5.34031e-04)
        (5.25000e+00, 1.58431e-04)
        (5.50000e+00, 3.99920e-05)
        (5.75000e+00, 9.45180e-06)
      };
      \addlegendentry{}
      \addlegendimage{empty legend}
      \addlegendentry{}
      \addlegendimage{empty legend}
      \addlegendentry{}
      \addlegendimage{empty legend}
      \addlegendentry{}

      \addlegendimage{empty legend}
      \addlegendentry[anchor=east]{SCF:}

      \addplot[very thick,color=cyan, mark=o] plot coordinates {
        (2.7500e+00, 8.4237e-01)
        (3.0000e+00, 6.4501e-01)
        (3.2500e+00, 3.9822e-01)
        (3.5000e+00, 1.9723e-01)
        (3.7500e+00, 7.4797e-02)
        (4.0000e+00, 2.3919e-02)
        (4.2500e+00, 5.9998e-03)
        (4.5000e+00, 1.5600e-03)
        (4.7500e+00, 3.7140e-04)
        (5.0000e+00, 8.5372e-05)
        (5.2500e+00, 1.7561e-05)
      };
      \addlegendentry{$T = 8$}

      \addplot[very thick,color=blue, mark=triangle] plot coordinates {
        (2.0000e+00, 9.9812e-01)
        (2.2500e+00, 9.8776e-01)
        (2.5000e+00, 9.4128e-01)
        (2.7500e+00, 8.1565e-01)
        (3.0000e+00, 6.0218e-01)
        (3.2500e+00, 3.5079e-01)
        (3.5000e+00, 1.6399e-01)
        (3.7500e+00, 5.6278e-02)
        (4.0000e+00, 1.6879e-02)
        (4.2500e+00, 4.0398e-03)
        (4.5000e+00, 1.1895e-03)
        (4.7500e+00, 2.7949e-04)
        (5.0000e+00, 6.9097e-05)
        (5.2500e+00, 1.6777e-05)
      };
      \addlegendentry{$T = 16$}

      \addlegendimage{empty legend}
      \addlegendentry{}
      \addlegendimage{empty legend}
      \addlegendentry{}

      \addlegendimage{empty legend}
      \addlegendentry[anchor=east]{SCL:}

      \addplot[very thick,dashed,color=magenta, mark=diamond] plot coordinates {
        (2.75000e+00, 7.93412e-01)
        (3.00000e+00, 5.61353e-01)
        (3.25000e+00, 3.15059e-01)
        (3.50000e+00, 1.46118e-01)
        (3.75000e+00, 5.25294e-02)
        (4.00000e+00, 1.65294e-02)
        (4.25000e+00, 4.00000e-03)
        (4.50000e+00, 1.19355e-03)
        (4.75000e+00, 2.43437e-04)
        (5.00000e+00, 6.45578e-05)
        (5.25000e+00, 1.20294e-05)
        (5.50000e+00, 2.49234e-06)
      };
      \addlegendentry{$L = 2$}

      \addplot[very thick,dashed,color=red, mark=x] plot coordinates {
        (2.25000e+00, 9.67941e-01)
        (2.50000e+00, 8.77000e-01)
        (2.75000e+00, 6.80882e-01)
        (3.00000e+00, 4.25235e-01)
        (3.25000e+00, 1.94706e-01)
        (3.50000e+00, 6.78824e-02)
        (3.75000e+00, 1.82353e-02)
        (4.00000e+00, 4.00000e-03)
        (4.25000e+00, 9.29204e-04)
        (4.50000e+00, 2.03593e-04)
        (4.75000e+00, 4.03551e-05)
        (5.00000e+00, 8.59476e-06)
      };
      \addlegendentry{$L = 4$}

      \addplot[very thick,dashed,color=orange, mark=asterisk] plot coordinates {
        (2.00000e+00, 9.91706e-01)
        (2.25000e+00, 9.52059e-01)
        (2.50000e+00, 8.29824e-01)
        (2.75000e+00, 5.99471e-01)
        (3.00000e+00, 3.36647e-01)
        (3.25000e+00, 1.29882e-01)
        (3.50000e+00, 3.87647e-02)
        (3.75000e+00, 8.04545e-03)
        (4.00000e+00, 1.65672e-03)
        (4.25000e+00, 3.45763e-04)
        (4.50000e+00, 8.48896e-05)
        (4.75000e+00, 1.74155e-05)
        (5.00000e+00, 4.72099e-06)
      };
      \addlegendentry{$L = 8$}

      \addplot[very thick,dashed,color=darkgreen, mark=square] plot coordinates {
        (2.00000e+00, 9.81158e-01)
        (2.25000e+00, 9.13088e-01)
        (2.50000e+00, 7.33123e-01)
        (2.75000e+00, 4.59526e-01)
        (3.00000e+00, 2.03825e-01)
        (3.25000e+00, 6.33860e-02)
        (3.50000e+00, 1.44211e-02)
        (3.75000e+00, 3.40351e-03)
        (4.00000e+00, 6.33540e-04)
        (4.25000e+00, 2.19780e-04)
        (4.50000e+00, 4.81696e-05)
        (4.75000e+00, 1.54250e-05)
      };
      \addlegendentry{$L = 32$}

    \end{semilogyaxis}
  \end{tikzpicture}
  \\
  \ref{ec-perf-legend}
	\caption{Error-correction performance comparison for a (1024,\,869) polar code decoded using three different algorithms. The SCL and SCF decoders use an 8-bit CRC.}
	\label{fig:ec-perf-cmp}
\end{figure}
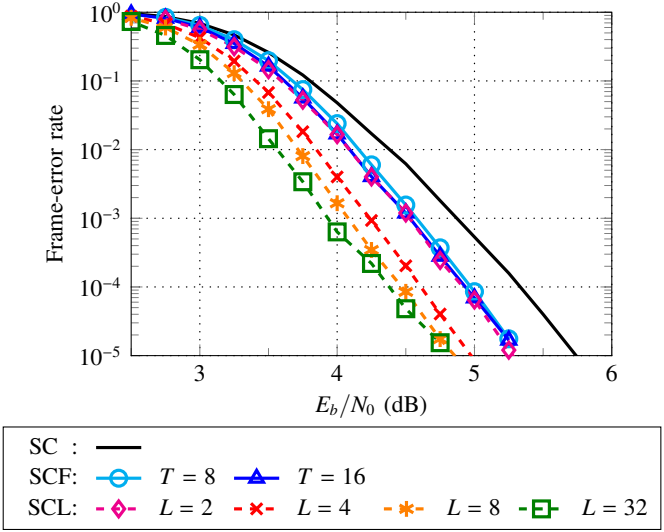

Fig.~\ref{fig:ec-perf-cmp} shows the error-correction performance of a (1024,\,869) polar code for three different decoding algorithms: \gls{sc}, \gls{scl}, and \gls{scf}. This particular code is used for comparison as this is also the code that is supported by the high-throughput fixed code-rate implementation of the \gls{sc} algorithm. These simulation results are for random codewords modulated with \gls{bpsk} and transmitted over an \gls{awgn} channel. For the \gls{scl} and \gls{scf} results, the polar code is concatenated with an 8-bit \gls{crc}, i.e., the number of information bits $k$ of the polar code is increased by 8 such that the code rate of the resulting system remains of $R=\nicefrac{869}{1024}$. The \gls{scf} algorithm was set to do a maximum number of trials $T$ of either 8 or 16. The list algorithm has a constrained list size $L$ of either 2, 4, or 32. From that figure, it can be seen that the \gls{sc} algorithm (black curve without markers) has the worst \gls{fer}. The \gls{scf} algorithm (blue curve with triangle markers and cyan curve with circle markers) offers a coding gain from approximately $0.35$\,dB to $0.4$\,dB at a \gls{fer} of $10^{-4}$ compared to the \gls{sc} algorithm. Both \gls{scf} curves are almost identical to the \gls{scl} results with $L=2$ (dashed-magenta curve with diamond markers). By increasing the list size $L$ to 4 (dashed-red curve with cross markers), the \gls{scl} algorithm improves the coding gain by $0.33$\,dB compared to the \gls{scf} results. Further increasing the list size $L$ to 32 (dashed-green curve with square markers) leads to a 0.31\,dB gain over $L=4$ up to a \gls{fer} of approximately $10^{-3}$ from which point the 8-bit \gls{crc} becomes too short to avoid collisions. This causes the gain to slowly degrate as the $\nicefrac{E_b}{N_0}$ ratio grows.

The gaps between these decoding algorithms depend on the parameters, however the order generally remains the same, i.e., \gls{sc} decoding will have the worst \gls{fer} of the three, while \gls{scl} decoding has the best one, and that of \gls{scf} decoding lies somewhere in between.

\begin{figure}[t]
	\centering
  \begin{tikzpicture}

    \pgfplotsset{
      grid style = {
        dash pattern = on 0.05mm off 1mm,
        line cap = round,
        black,
        line width = 0.5pt
      },
      label style = {font=\fontsize{9pt}{7.2}\selectfont},
      tick label style = {font=\fontsize{9pt}{7.2}\selectfont}
    }

    \begin{semilogyaxis}[%
      xlabel=$E_b/N_0$ (dB),xtick={0,1,2,...,6.0},%
      xlabel style={yshift=0.4em},%
      minor x tick num={1},
      xmin=0,xmax=6,%
      ymin=1e-5,ymax=1e0,%
      ylabel=Frame-error rate, ylabel style={yshift=-1.0em},%
      width=0.535\columnwidth, height=6.235cm, grid=major,%
      legend style={
        anchor={center},
        cells={anchor=west},
        column sep=2mm,
        font=\fontsize{9pt}{7.2}\selectfont,
        mark size=3.0pt
      },
      legend columns=6,
      legend to name=flip-ec-perf-legend,
      mark size=3.0pt,
      mark options=solid]
      
      \addlegendimage{empty legend}
      \addlegendentry[anchor=east]{$R$:}

      \addplot[very thick,color=red, mark=x] plot coordinates {
        (0.0000e+00, 7.8957e-01)
        (2.5000e-01, 6.4657e-01)
        (5.0000e-01, 4.8090e-01)
        (7.5000e-01, 3.2475e-01)
        (1.0000e+00, 1.8963e-01)
        (1.2500e+00, 9.7756e-02)
        (1.5000e+00, 4.4718e-02)
        (1.7500e+00, 2.0359e-02)
        (2.0000e+00, 6.9997e-03)
        (2.2500e+00, 2.1887e-03)
        (2.5000e+00, 8.0738e-04)
        (2.7500e+00, 2.2468e-04)
        (3.0000e+00, 5.4083e-05)
        (3.2500e+00, 1.0297e-05)
      };
      \addlegendentry{$\nicefrac{1}{4}$}

      \addplot[very thick,color=blue, mark=triangle] plot coordinates {
        (1.0000e+00, 6.8589e-01)
        (1.2500e+00, 4.7370e-01)
        (1.5000e+00, 2.6947e-01)
        (1.7500e+00, 1.2811e-01)
        (2.0000e+00, 4.7758e-02)
        (2.2500e+00, 1.5879e-02)
        (2.5000e+00, 4.5198e-03)
        (2.7500e+00, 1.0373e-03)
        (3.0000e+00, 2.2038e-04)
        (3.2500e+00, 5.0864e-05)
        (3.5000e+00, 9.6838e-06)
      };
      \addlegendentry{$\nicefrac{1}{2}$}

      \addplot[very thick,color=black] plot coordinates {
        (1.5000e+00, 8.9308e-01)
        (1.7500e+00, 7.3601e-01)
        (2.0000e+00, 5.1314e-01)
        (2.2500e+00, 2.9183e-01)
        (2.5000e+00, 1.3167e-01)
        (2.7500e+00, 4.4678e-02)
        (3.0000e+00, 1.4239e-02)
        (3.2500e+00, 3.5569e-03)
        (3.5000e+00, 6.8452e-04)
        (3.7500e+00, 1.0610e-04)
        (4.0000e+00, 2.1086e-05)
        (4.2500e+00, 2.6931e-06)
      };
      \addlegendentry{$\nicefrac{2}{3}$}

      \addplot[very thick,color=cyan, mark=o] plot coordinates {
        (2.0000e+00, 8.6505e-01)
        (2.2500e+00, 6.8837e-01)
        (2.5000e+00, 4.4570e-01)
        (2.7500e+00, 2.3255e-01)
        (3.0000e+00, 9.4316e-02)
        (3.2500e+00, 3.0599e-02)
        (3.5000e+00, 8.5597e-03)
        (3.7500e+00, 2.2301e-03)
        (4.0000e+00, 3.8726e-04)
        (4.2500e+00, 6.7519e-05)
        (4.5000e+00, 1.2556e-05)
      };
      \addlegendentry{$\nicefrac{3}{4}$}

      \addplot[very thick,color=darkgreen, mark=diamond] plot coordinates {
        (2.7500e+00, 8.4237e-01)
        (3.0000e+00, 6.4501e-01)
        (3.2500e+00, 3.9822e-01)
        (3.5000e+00, 1.9723e-01)
        (3.7500e+00, 7.4797e-02)
        (4.0000e+00, 2.3919e-02)
        (4.2500e+00, 5.9998e-03)
        (4.5000e+00, 1.5600e-03)
        (4.7500e+00, 3.7140e-04)
        (5.0000e+00, 8.5372e-05)
        (5.2500e+00, 1.7561e-05)
      };
      \addlegendentry{$\nicefrac{5}{6}$}

    \end{semilogyaxis}
  \end{tikzpicture}\hspace{-5pt}
  \begin{tikzpicture}

    \pgfplotsset{
      grid style = {
        dash pattern = on 0.05mm off 1mm,
        line cap = round,
        black,
        line width = 0.5pt
      },
      label style = {font=\fontsize{9pt}{7.2}\selectfont},
      tick label style = {font=\fontsize{9pt}{7.2}\selectfont}
    }

    \begin{semilogyaxis}[%
      xlabel=$E_b/N_0$ (dB),xtick={0,1,2,...,6.0},%
      xlabel style={yshift=0.4em},%
      minor x tick num={1},
      xmin=-0.5,xmax=5.5,%
      ymin=1e-5,ymax=1e0,%
      ylabel=Frame-error rate, ylabel style={yshift=-1.0em},%
      width=0.535\columnwidth, height=6.235cm, grid=major,%
      mark size=3.0pt,
      mark options=solid]
      
      \addplot[very thick,dashed,color=red, mark=x] plot coordinates {
        (-5.00000e-01, 8.18404e-01)
        (-2.50000e-01, 6.76105e-01)
        (0.00000e+00, 5.07228e-01)
        (2.50000e-01, 3.63702e-01)
        (5.00000e-01, 2.11193e-01)
        (7.50000e-01, 1.06754e-01)
        (1.00000e+00, 4.43158e-02)
        (1.25000e+00, 1.70351e-02)
        (1.50000e+00, 5.45614e-03)
        (1.75000e+00, 1.44304e-03)
        (2.00000e+00, 4.04000e-04)
        (2.25000e+00, 1.10142e-04)
        (2.50000e+00, 2.96648e-05)
        (2.75000e+00, 8.09879e-06)
      };

      \addplot[very thick,dashed,color=blue, mark=triangle] plot coordinates {
        (2.50000e-01, 9.22754e-01)
        (5.00000e-01, 7.90228e-01)
        (7.50000e-01, 5.83298e-01)
        (1.00000e+00, 3.86404e-01)
        (1.25000e+00, 1.88491e-01)
        (1.50000e+00, 7.43860e-02)
        (1.75000e+00, 2.33509e-02)
        (2.00000e+00, 6.17544e-03)
        (2.25000e+00, 1.50000e-03)
        (2.50000e+00, 2.40566e-04)
        (2.75000e+00, 6.20732e-05)
        (3.00000e+00, 1.57011e-05)
        (3.25000e+00, 3.84586e-06)
      };

      \addplot[very thick,dashed,color=black] plot coordinates {
        (1.00000e+00, 9.64333e-01)
        (1.25000e+00, 8.72053e-01)
        (1.50000e+00, 6.84930e-01)
        (1.75000e+00, 4.30526e-01)
        (2.00000e+00, 2.05702e-01)
        (2.25000e+00, 7.23509e-02)
        (2.50000e+00, 1.99298e-02)
        (2.75000e+00, 4.78947e-03)
        (3.00000e+00, 7.55396e-04)
        (3.25000e+00, 1.26092e-04)
        (3.50000e+00, 2.76855e-05)
        (3.75000e+00, 4.31648e-06)
      };

      \addplot[very thick,dashed,color=cyan, mark=o] plot coordinates {
        (1.50000e+00, 9.66965e-01)
        (1.75000e+00, 8.76281e-01)
        (2.00000e+00, 6.82281e-01)
        (2.25000e+00, 4.20912e-01)
        (2.50000e+00, 1.95702e-01)
        (2.75000e+00, 6.66667e-02)
        (3.00000e+00, 1.74737e-02)
        (3.25000e+00, 3.87719e-03)
        (3.50000e+00, 4.83871e-04)
        (3.75000e+00, 1.16494e-04)
        (4.00000e+00, 1.99045e-05)
        (4.25000e+00, 3.51605e-06)
      };

      \addplot[very thick,dashed,color=darkgreen, mark=diamond] plot coordinates {
        (2.25000e+00, 9.67941e-01)
        (2.50000e+00, 8.77000e-01)
        (2.75000e+00, 6.80882e-01)
        (3.00000e+00, 4.25235e-01)
        (3.25000e+00, 1.94706e-01)
        (3.50000e+00, 6.78824e-02)
        (3.75000e+00, 1.82353e-02)
        (4.00000e+00, 4.00000e-03)
        (4.25000e+00, 9.29204e-04)
        (4.50000e+00, 2.03593e-04)
        (4.75000e+00, 4.03551e-05)
        (5.00000e+00, 8.59476e-06)
      };

    \end{semilogyaxis}
  \end{tikzpicture}
  \\
  \ref{flip-ec-perf-legend}
	\caption{Error-correction performance comparison for polar codes of blocklength $N=1024$ with a variable code rate $R$ decoded using either the SCF algorithm (left, solid curves) or the SCL algorithm (right, dashed curves). The SCF maximum number of trials $T=8$, the SCL list size $L=4$; results are for an 8-bit CRC.}
	\label{fig:flip-ec-perf-cmp}\vspace{-8pt}
\end{figure}
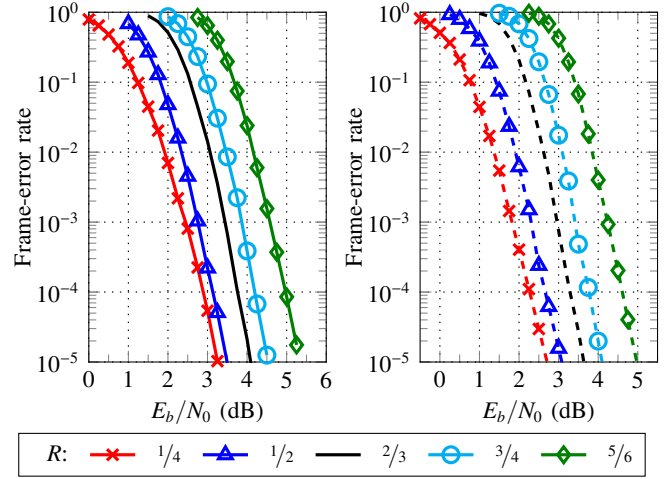

Fig.~\ref{fig:flip-ec-perf-cmp} shows the error-correction performance of polar codes of blocklength $N=1024$, for various code rates, under \gls{scf} and \gls{scl} decoding. These \gls{fer} results are included for reference as these are the codes used for the measurement results presented in Section~\ref{sec:results}.

\vspace{-3pt}
\section{\polarbear Architecture}\label{sec:chip_architecture}

\begin{figure}[t]
	\centering
  \includegraphics[width=.7\columnwidth]{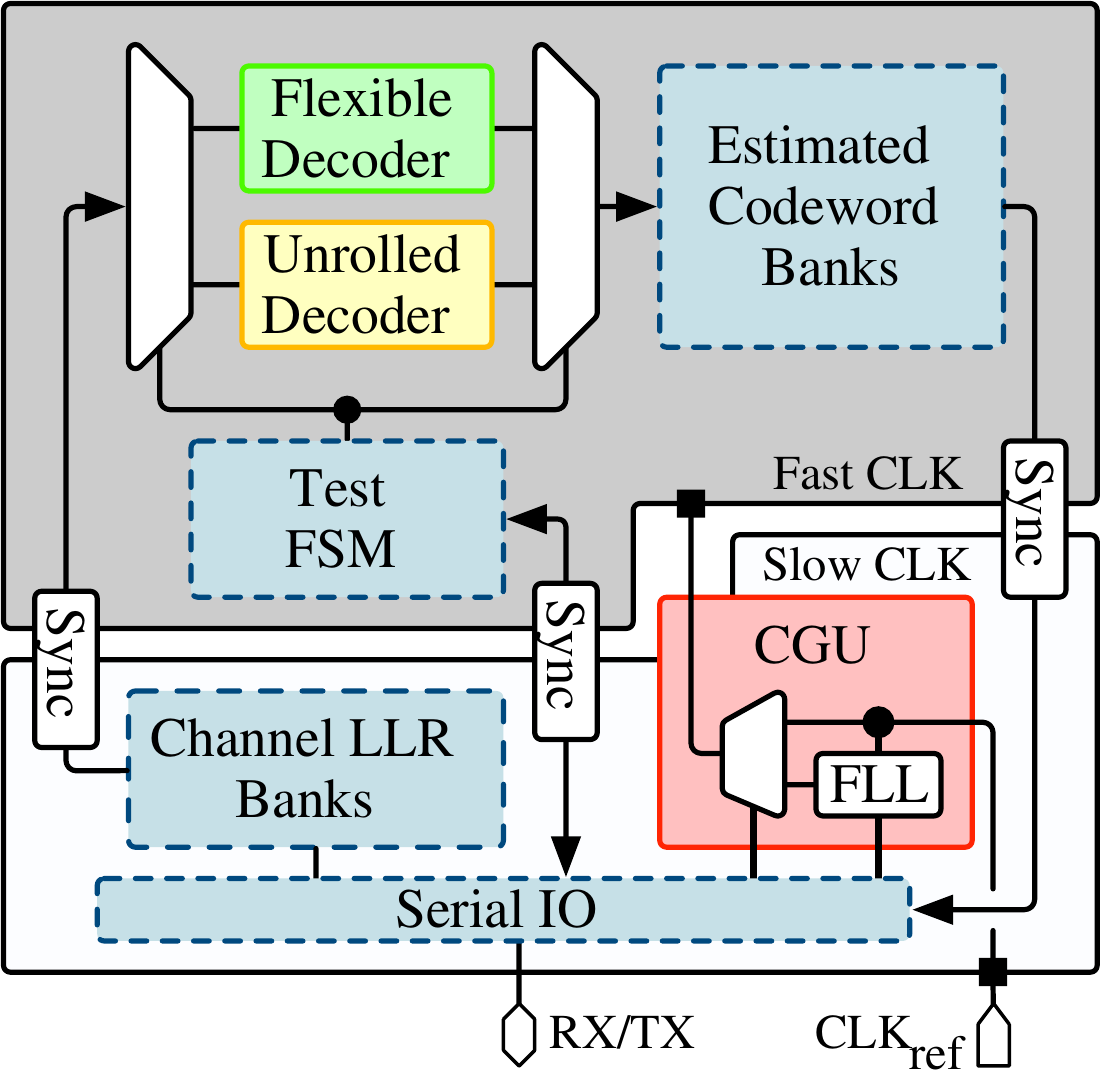}
	\caption{Simplified overview of the \polarbear architecture. The Test-Controller Unit (TCU) is composed of the modules highlighted in blue with a dashed outline.}
	\label{fig:polarbear-arch}
\end{figure}

Fig.~\ref{fig:polarbear-arch} shows an overview of the \polarbear chip architecture. \polarbear comprises four main units: the flexible decoder, in green, the unrolled decoder, in yellow, the \gls{cgu}, in red, and the \gls{tcu}, made of multiple modules, all illustrated in blue with a dashed outline.
Both decoders represent channel and internal soft values as quantized \glspl{llr} in the 2's complement format. We denote quantization as $Q_i$.$Q_c$, where $Q_c$ is the total number of bits to store a channel \gls{llr} and $Q_i$ is the number of bits used to store an internal \gls{llr}. Both decoders have quantization parameters that can be modified at the time of synthesis.

There are multiple power domains on the chip, supplied through distinct pins.  This allows to precisely measure the current drawn by each of the two decoders.

There are two clock domains on the chip. One is slower---typically around 20\,MHz---and is used as a reference clock for the \gls{cgu} as well as by some of the \gls{tcu} modules. The faster clock is used by the decoders, the test \gls{fsm}, and to read from the channel-\gls{llr} banks and to write to estimated-codeword banks. A serial interface, which is part of the \gls{tcu}, provides the means to communicate with the \polarbear chip from the outside world. Section~\ref{sec:arch:tcu} provides a more detailed description of the \gls{tcu}.

\begin{figure*}
	\centering
  \includegraphics[width=.95\textwidth]{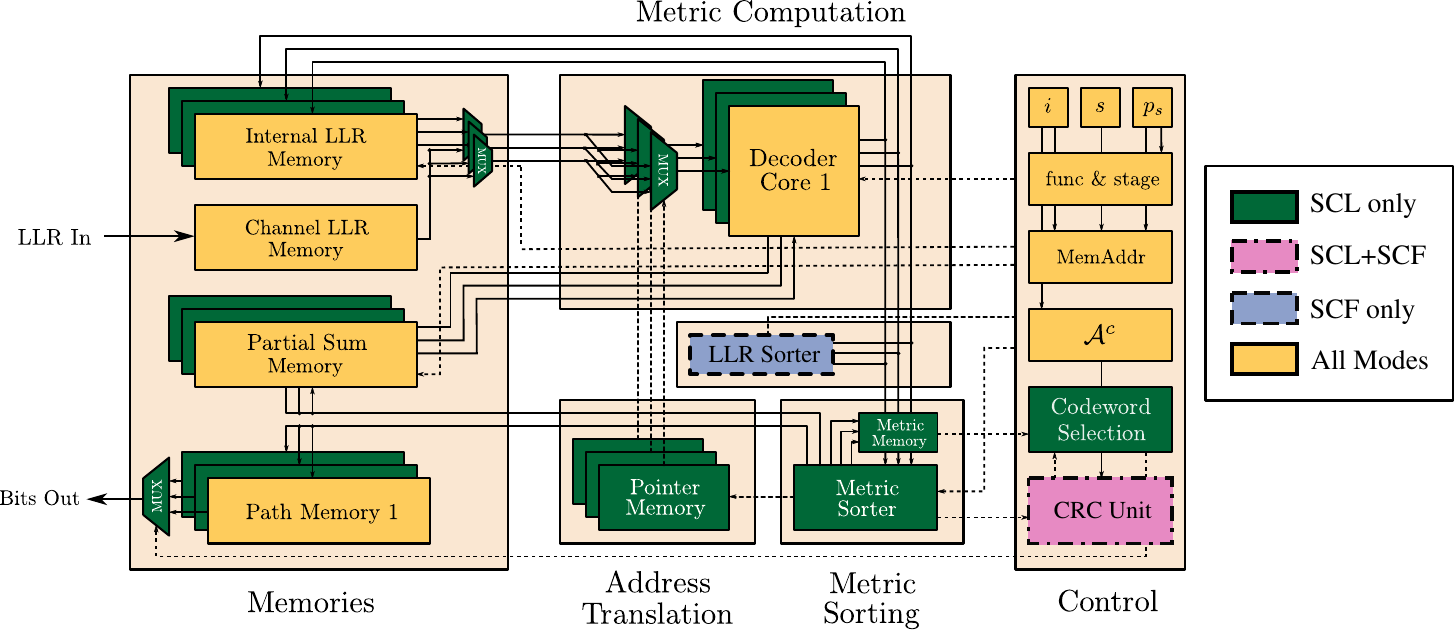}
	\caption{Flexible-decoder architecture. In SCL decoder mode, all modules but the \gls{llr} sorter unit are used. The modules used in SCF decoder mode are colored in orange, in purple with a dashed-dotted outline, and in blue with a dashed outline. The SC mode only uses the modules colored in orange.}
	\label{fig:flexible-arch}\vspace{-8pt}
\end{figure*}

\subsection{Flexible Decoder}
The flexible decoder supports all three decoding algorithms described in the previous section, i.e., \gls{sc} decoding, \gls{scf} decoding, and \gls{scl} decoding.
This decoder also supports decoding of polar codes of any rate for a given blocklength $N$, various list sizes ranging from $L=2$ up to a maximum list size $L = L_{\max}$ for \gls{scl} decoding, and a configurable maximum number of decoding trials $T_{\max}$ for the \gls{scf} decoding algorithm. In this architecture, $L_{\max}$ decoder cores are instantiated. Moreover, the \gls{crc} unit supports various \gls{crc} lengths in order to implement CRC-aided \gls{scl} decoding, and \gls{scf} decoding. The \gls{crc} length can be selected during runtime.

\subsubsection*{Architecture Overview} 
An overview of the flexible decoder architecture is presented in Fig.~\ref{fig:flexible-arch} along with a legend explaining which components are used for the different supported decoding modes.

More specifically, the decoder contains one memory bank for the channel \glspl{llr} and $L_{\max}$ memory banks for the internal \glspl{llr} and the partial sums. Moreover, there are $L_{\max}$ memory banks that form the path memory, which is used to store the paths taken along the decoding tree, which correspond to candidate codewords. We note that, for SC decoding of a non-systematic polar code, it is not strictly necessary to use the path memory as there is only a single candidate codeword which can be output serially as decoding proceeds. However, in our decoder architecture the single candidate codeword is stored even for \gls{sc} decoding, as this enables the decoder to also decode systematic polar codes when used in conjuction with a re-encoding block to obtain the information bits. There are $L_{\max}$ decoder cores which implement the basic update rules for \gls{sc} decoding. A single decoder core is used during \gls{sc} and \gls{scf} decoding, while up to $L_{\max}$ decoder cores are used during \gls{scl} decoding, depending on the employed list size. The flexible decoder also contains two sorting units, namely the path-metric sorter (identified as ``metric sorter'' for short, in Fig.~\ref{fig:flexible-arch}) and the \gls{llr} sorter, which are used during \gls{scl} and \gls{scf} decoding, respectively. The path-metric sorter is used to identify the $L$ most reliable decoding paths out of the $2L$ candidate decoding paths that are produced every time the SCL decoder encounters an information bit. We use a pruned radix-$2L$ sorter in order to sort the path metric as it is the fastest sorter for $L_{\max} = 4$~\cite{Balatsoukas_ISCAS_2015}. The \gls{llr} sorter, on the other hand, is used in order to identify the $T-1$ information bits with the smallest decision-\gls{llr} absolute values, which correspond to the $T-1$ least reliable decisions. The \gls{llr} sorter architecture is described in more detail in Section~\ref{sec:scf}. Finally, the decoder contains a pointer memory, which implements the low-complexity state copying mechanism for \gls{scl} decoding as described in detail in~\cite{Balatsoukas_TCASII_2014}, as well as a controller which is responsible for the generation of all control signals and for the calculation of the \gls{crc} for \gls{scl} and \gls{scf} decoding.

The set of frozen-bit locations $\mathcal{A}^c$ is derived from a $N$-bit wide binary vector provided at the input, where a one or a zero indicate that the location corresponds to a frozen bit or an information bit, respectively.

\subsubsection*{Latency Saving Technique} Since the values of frozen bits are known a priori at the receiver, no \gls{llr} computations are in fact necessary until the first non-frozen bit is reached during the SC decoding process. This observation is exploited in our decoder in order to directly start decoding from the first information bit and reduce the decoding latency. Note that this latency reduction technique can be seen as partial application of the SSC algorithm~\cite{Alamdar-Yazdi2011}, with the important advantage that it is applicable verbatim to \gls{scl} decoding, as the first path fork only occurs at the first information-bit location.

In the following sections, we provide more details on each of the different decoding modes.

\subsubsection{SCL Mode}
The flexible decoder implements the \gls{scl} decoding algorithm as briefly reviewed in Section~\ref{sec:bg:scl} and as more thoroughly described in \cite{Balatsoukas_TSP_2015}.
The \gls{scl} decoder implementation requires all modules illustrated in Fig.~\ref{fig:flexible-arch}, except for the \gls{llr} sorting unit that is only used by the \gls{scf} decoder. The \gls{crc} calculations take place alongside the decoding process, as the information bits become available one by one, and thus do not incur any additional latency. Moreover, this characteristic enables a very compact serial implementation of the \gls{crc} units, rendering their size negligible.

\subsubsection{SC Mode}
The flexible decoder also implements a slightly improved version of the original \gls{sc} algorithm~\cite{Arikan2009}. The improvement consists in the latency reduction technique described above, i.e., a priori knowledge of the first information-bit location allows the algorithm to skip the unnecessary calculations that would otherwise mandate the \gls{sc} algorithm to visit frozen bit locations.

As illustrated in Fig.~\ref{fig:flexible-arch}, the \gls{sc} decoder mode only uses one of the $L_{\max}$ decoder cores. Moreover, the \gls{sc} mode uses only one of the internal-\gls{llr}-memory banks, one of the partial-sum-memory banks, and one of the path-memory banks. For \gls{sc} operation both the path-metric sorting unit and the \gls{llr} sorting unit are bypassed completely.

\subsubsection{SCF Mode}\label{sec:scf}
The flexible decoder also implements the \gls{scf} decoding algorithm as proposed in \cite{Afisiadis2014}, and as briefly described in Section~\ref{sec:bg:scf}. Similarly to the \gls{sc} decoder, the \gls{scf} decoder mode only uses one of the $L_{\max}$ decoder cores, a single internal-\gls{llr}-memory bank, a single partial-sum-memory bank, and a single path-memory bank. These components are illustrated in orange (labeled as ``All Modes'' in the legend) in Fig.~\ref{fig:flexible-arch}. In addition to the hardware required for \gls{sc} decoding, the \gls{scf} decoder uses the \gls{crc} unit, colored in purple with a dashed-dotted outline, and a dedicated \gls{llr} sorter, colored in blue with a dashed outline, that identifies the $T-1$ least reliable bit-decisions during the first decoding attempt, i.e., the bit-decisions that had the $T-1$ smallest absolute \gls{llr} values.
 
\begin{figure}
	\centering
  \includegraphics[width=.8\columnwidth]{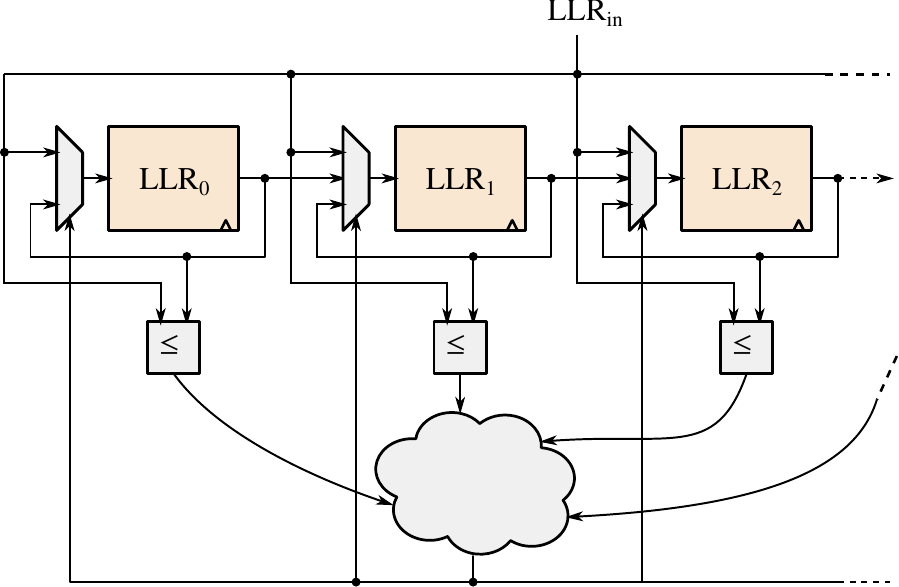}
	\caption{Insertion sorter used in the SCF decoder to identify the $T-1$ least reliable bit-decisions.}
	\label{fig:llr-sorter}\vspace{-8pt}
\end{figure}

Since the decision \glspl{llr} that need to be sorted become available at a rate of at most one \gls{llr} per clock cycle, an insertion sorter was selected to implement the \gls{llr} sorter. The insertion sorter can be fully parallelized in order to sort each \gls{llr} in a single clock cycle. More specifically, each decision \gls{llr} is compared in parallel with all $T-1$ existing (and already sorted) least reliable decision \glspl{llr} which are stored in registers. Using the result of these comparisons, it is straightforward to decide whether the new \gls{llr} should be stored and to identify the location in which it should be inserted. Insertion can then be performed efficiently in a single clock cycle by shifting the content of the registers that are after the insertion position by one position, discarding the \gls{llr} at position $T-1$ in the process, and writing the new \gls{llr} value in its corresponding position, while keeping the remaining contents at their place. We note that the registers containing the $T-1$ least reliable decision \glspl{llr} are initialized to the maximum possible absolute \gls{llr} value when decoding starts. A high level block diagram of the sorter is presented in Fig.~\ref{fig:llr-sorter}.

\subsection{Decoding Latency and Throughput of the Flexible Decoder}
Since all three algorithms implemented by the flexible decoder are based on \gls{sc} decoding, their decoding latency is largely dictated by the decoding latency of the underlying \gls{sc} hardware decoder. More specifically, the time required by the \gls{sc} decoding algorithm to generate an estimated codeword, measured in \glspl{cc}, can be expressed as:
\begin{equation}\label{eqn:latency:sc}
  \mathcal{L}_{\text{SC}} = 2N+\frac{N}{64}\log_2\left(\frac{N}{256}\right) -
  \sum\limits_{i=0}^{\log_2N} \left\lfloor \frac{b}{2^i}\right\rfloor \left\lceil \frac{2^i}{64} \right\rceil,
\end{equation}
where $N$ is the polar-code blocklength, and $b$ is the location of the first information bit. The two left-hand-side terms correspond to the latency of a semi-parallel \gls{sc} decoder implementation~\cite{Leroux2013}, where $P=64$. The right-hand-side term is a correction term that stems from the polar-code-specific simplifications described earlier, a contribution of this work.

The \gls{scl} algorithm performs some additional steps compared to the \gls{sc} algorithm. In particular, the metric sorting step involved in \gls{scl} decoding cannot be performed in parallel with the \gls{llr} computations and thus increases the latency of the \gls{scl} decoder with respect to that of the \gls{sc} decoder. More specifically, the latency of \gls{scl} decoding depends on the code rate and on the distribution of frozen-bit \textit{clusters} in the polar code. Let us partition $\mathcal{A}^{\text{c}}$ as $\mathcal{A}^{\text{c}} = \bigcup_{j=1}^{\mathcal{F}_C} \mathcal{A}^{\text{c}}_j$ such that:
\begin{itemize}
	\item[(i)] $\mathcal{A}^{\text{c}}_j \cap \mathcal{A}^{\text{c}}_{j'} = \emptyset$ if $j \ne j'$,
	\item[(ii)] for every $j$, $\mathcal{A}^{\text{c}}_j$ is a contiguous subset of $\{0,\hdots,N-1\}$,
	\item[(iii)] for every pair $j \ne j'$, $\mathcal{A}^{\text{c}}_j \cup \mathcal{A}^{\text{c}}_{j'}$ is \emph{not} a contiguous subset of $\{0,\hdots,N-1\}$.
\end{itemize}
Then, each $\mathcal{A}^{\text{c}}_j$ is a frozen-bit cluster and $\mathcal{F}_C$ is the total number of frozen-bit clusters in a polar code.

Using the above definition of a frozen-bit cluster, the latency of the \gls{scl} decoding algorithm is given by:
\begin{equation}\label{eqn:latency:scl}
  \mathcal{L}_{\text{SCL}} = \mathcal{L}_{\text{SC}}+\mathcal{L}_{\text{sort}},
\end{equation}
where $\mathcal{L}_{\text{SC}}$ is the latency of the \gls{sc} decoder as defined in~\eqref{eqn:latency:sc} and $\mathcal{L}_{\text{sort}}$ is the latency incurred by the sorting steps defined as~\cite{Balatsoukas_TSP_2015}:
\begin{equation}\label{eqn:latency:scl:sort}
  \mathcal{L}_{\text{sort}} = k + \mathcal{F}_C,
\end{equation}
where $k$ is the number of information bits and $\mathcal{F}_C$ is the number of frozen-bit \textit{clusters}. Similarly to the right-hand-side term of \eqref{eqn:latency:sc}, $\mathcal{F}_C$ is also polar-code specific. 

Contrary to \gls{sc} and \gls{scl} decoding, \gls{scf} decoding has a variable runtime that depends on the number of performed decoding attempts. The worst-case latency of the \gls{scf} decoding algorithm can be expressed as:
\begin{equation}\label{eqn:latency:scf}
  \mathcal{L}_{\text{SCF}} = T\mathcal{L}_{\text{SC}},
\end{equation}
where $T$ is the maximum number of trials, and $\mathcal{L}_{\text{SC}}$ is the latency of the \gls{sc} decoder as defined in~\eqref{eqn:latency:sc}. It is noteworthy that, as will be shown in the sequel, for the \gls{fer} values of interest the average latency of \gls{scf} decoding is very close to that of standard \gls{sc} decoding.

Since only a single codeword is decoded at any given time by the flexible decoder, the decoding throughput can be directly calculated from the decoding latency. Thus, the coded throughput of the flexible decoder is given by:
\begin{equation}
  \mathcal{T_{\text{x}}} = \frac{Nf_{\text{clk}}}{\mathcal{L}_{\text{x}}}\;\text{bps},
\end{equation}
where $\text{x} \in \{\text{SC}, \text{SCF},\text{SCL}\}$.

\subsection{Fully-Unrolled Partially-Pipelined SC Decoder}
The \gls{sc} decoder implementation is optimized for speed and energy efficiency at the expense of flexibility and error-correction performance (compared to the \gls{scl} and \gls{scf} decoding algorithms), and is based on the fast-SSC algorithm~\cite{Sarkis_JSAC_2014} and on a fully-unrolled partially-pipelined architecture for a polar decoder as presented in~\cite{Giard_TCAS_2016}.

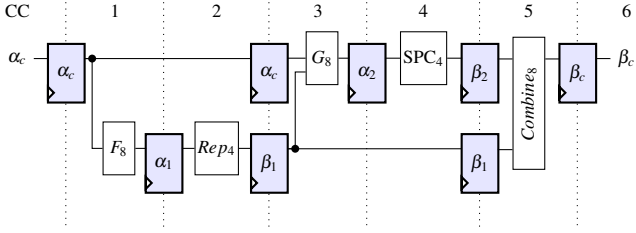
\begin{figure}[t]
	\centering
  \begin{tikzpicture}[font=\scriptsize,inner sep=1pt, minimum width=1.2em]

    \definecolor{deepgreen}{RGB}{8, 130, 25}

    \tikzset{
      branch/.style={fill,shape=circle,minimum size=3pt,inner sep=0pt},
      block/.style={draw, rectangle, minimum height=2em},
      spc/.style={draw, rectangle, minimum height=2em},
      rep/.style={draw, rectangle, minimum height=2em},
      comb/.style={draw, rectangle, minimum height=5em}
    }

    \node (ac) at (0.2,0) {$\alpha_c$};
    \node at (0.2, 0.63) (cc) {CC};

    \node[shape=reg] at ($(ac)+(0.65,-0.195)$) (REG1) {$\alpha_c$};

    \phantom{\node[shape=reg] at ($(REG1)+(1.3,0)$) (REG2) {$\alpha_c$};}
    \node[block] at ($(REG1)+(0.7,-1.0)$) (F1) {$F_8$};
    \node[shape=reg] at ($(F1)+(0.6,-0.2)$) (REG4) {$\alpha_1$};

    \node[shape=reg] at ($(REG2)+(1.4,0)$) (REG3) {$\alpha_c$};
    \node[rep] at ($(REG4)+(0.7,0.2)$) (Rep1) {$Rep_4$};
    \node[shape=reg] at ($(Rep1)+(0.7,-0.2)$) (REG5) {$\beta_1$};

    \node[block] at ($(REG3)+(0.7,0.2)$) (G1) {$G_8$};
    \node[shape=reg] at ($(REG3)+(1.3,0)$) (REG6) {$\alpha_2$};
    \phantom{\node[shape=reg] at ($(REG5)+(1.3,0)$) (REG7) {$\beta_1$};}

    \node[spc] at ($(REG6)+(0.75,0.2)$) (SPC1) {SPC$_4$};
    \node[shape=reg] at ($(SPC1)+(0.75,-0.2)$) (REG8) {$\beta_2$};
    \node[shape=reg] at ($(REG7)+(1.5,0)$) (REG9) {$\beta_1$};

    \node[comb] at ($(REG8)+(0.65,-0.4)$) (Comb1) {\rotatebox{90}{$Combine_8$}};
    \node[shape=reg] at ($(Comb1)+(0.65,0.4)$) (REG10) {$\beta_c$};

    \node (bc) at ($(REG10)+(0.65,0.195)$) {$\beta_c$};
    \draw let \p1 = (bc), \p2 = (cc) in node[coordinate] at (\x1, 0.63) (cc6) {6};
    \node at (cc6) {6};

    \draw[-] (ac.east) -- (REG1.D);

    \draw[dotted] ($(REG1.Q)+(-0.25,0.75)$) -- ($(REG1.Q)+(-0.25,0.2)$) ($(REG1.Q)+(-0.25,-0.6)$) -- ($(REG1.Q)+(-0.25,-2.25)$);

    \draw[-] (REG1.Q) -- (REG3.D);
    \draw[-] (REG1.Q) -- ++(0.1,0) node[branch] {} -- ++(0,-0.45) |- (F1);
    \draw[-] (F1) -| (REG4.D);

    \draw[dotted] ($(REG2.Q)+(-0.25,0.75)$) -- ($(REG4.Q)+(-0.25,0.2)$) ($(REG4.Q)+(-0.25,-0.6)$) -- ($(REG2.Q)+(-0.25,-2.25)$);

    \draw[-] (REG4.Q) |- (Rep1) -| (REG5.D);

    \draw[dotted] ($(REG3.Q)+(-0.25,0.75)$) -- ($(REG3.Q)+(-0.25,0.2)$) ($(REG3.Q)+(-0.25,-0.6)$) -- ($(REG5.Q)+(-0.25,0.2)$) ($(REG5.Q)+(-0.25,-0.6)$) -- ($(REG3.Q)+(-0.25,-2.25)$);

    \draw[-] (REG3.D) (REG3.Q) |- (G1) -| (REG6.D);
    \draw[-] (REG5.Q) -- ++(0.1,0) node[branch] {} |- ([yshift=-3mm]G1);
    \draw[-] (REG5.Q) -- (REG9.D);

    \draw[dotted] ($(REG6.Q)+(-0.25,0.75)$) -- ($(REG6.Q)+(-0.25,0.2)$) ($(REG6.Q)+(-0.25,-0.6)$) -- ($(REG6.Q)+(-0.25,-2.25)$);

    \draw[-] (REG6.Q) |- (SPC1) -| (REG8.D);

    \draw[dotted] ($(REG8.Q)+(-0.25,0.75)$) -- ($(REG8.Q)+(-0.25,0.2)$) ($(REG8.Q)+(-0.25,-0.6)$) -- ($(REG9.Q)+(-0.25,0.2)$) ($(REG9.Q)+(-0.25,-0.6)$) -- ($(REG8.Q)+(-0.25,-2.25)$);

    \draw[-] (REG8.Q) |- ([yshift=11mm]Comb1);
    \draw[-] (REG9.Q) |- ([yshift=-11.5mm]Comb1);

    \draw[-] (REG10.D) |- ([yshift=11mm]Comb1);

    \draw[dotted] ($(REG10.Q)+(-0.25,0.75)$) -- ($(REG10.Q)+(-0.25,0.2)$) ($(REG10.Q)+(-0.25,-0.6)$) -- ($(REG10.Q)+(-0.25,-2.25)$);

    \draw[-] (bc.west) |- (REG10.Q);

    \draw let \p1 = (REG1), \p2 = (cc) in node[coordinate] at (\x1, \y2) (cc0) {0};
    \draw let \p1 = (REG4), \p2 = (cc) in node[coordinate] at (\x1, \y2) (cc1) {1};
    \draw let \p1 = (REG3), \p2 = (cc) in node[coordinate] at (\x1, \y2) (cc2) {2};
    \draw let \p1 = (REG6), \p2 = (cc) in node[coordinate] at (\x1, \y2) (cc3) {3};
    \draw let \p1 = (REG8), \p2 = (cc) in node[coordinate] at (\x1, \y2) (cc4) {4};
    \draw let \p1 = (REG10), \p2 = (cc) in node[coordinate] at (\x1, \y2) (cc5) {5};

    \node at ($(cc0)!0.5!(cc1)$) {1};
    \node at ($(cc1)!0.5!(cc2)$) {2};
    \node at ($(cc2)!0.5!(cc3)$) {3};
    \node at ($(cc3)!0.5!(cc4)$) {4};
    \node at ($(cc4)!0.5!(cc5)$) {5};

  \end{tikzpicture}
	\caption{Fully-unrolled partially-pipelined SC decoder architecture example for a (8, 4) polar code, where the initiation interval $\mathcal{I}$ equals 2. Clock gates and signals omitted for clarity.}
	\label{fig:unrolledsc-arch}
\end{figure}

Fig.~\ref{fig:unrolledsc-arch} illustrates an example of a fully-unrolled partially-pipelined \gls{sc} decoder for the (8,\,4) polar code represented as a graph in Fig.~\ref{fig:pc8-enc}. Partial pipelining, as opposed to deep pipelining, allows to reduce the required area, at the cost of reducing the throughput, by removing redundant shimming registers in parts of the pipeline where data remains unchanged over multiple clock cycles~\cite{Giard_TCAS_2016}. In this example the initiation interval is $\mathcal{I}=2$, meaning that, at every second clock cycle, a new frame can be fed into the decoder and a new codeword is estimated. In Fig.~\ref{fig:unrolledsc-arch}, registers are shown in light blue, where $\alpha$ and $\beta$ registers are for \glspl{llr} and bit-vector estimates, respectively. The blocks in white, marked $F$, $G$, $Combine$, $Rep$, and SPC, correspond to functions of the fast-SSC algorithm, and the subscript indicates their respective width.
Data flows from left to right with very little control logic.

The latency of our unrolled decoder is polar-code specific as it depends on the distribution of the frozen bit locations~\cite{Sarkis_JSAC_2014}, but it is by nature significantly smaller than $\mathcal{L}_{\text{SC}}$. An example of that difference is given in Table~\ref{tab:latency}. The coded throughput of a fully-unrolled decoder does not depend on the distribution of the frozen bit locations and is given by:
\begin{equation}
  \mathcal{T_{\text{U-SC}}} = \frac{Nf_{\text{clk}}}{\mathcal{I}}\;\text{bps},
\end{equation}
where $f_{\text{clk}}$ is the clock frequency of the decoder.

\subsection{Clock-Generation and Test-Controller Units}\label{sec:arch:tcu}
The \gls{cgu}, highlighted in red in Fig.~\ref{fig:polarbear-arch}, produces a fast clock from a reference clock by using a flexible configurable \gls{fll}~\cite{Miro2014}. The \gls{cgu} has its own supply $\text{V}_{\text{CGU}}$ such that its energy consumption does not affect the decoder measurements.

The \gls{tcu} is the interface to the decoders and the \gls{fll}. The majority of its area consists of memory, which is implemented using registers. More specifically, there are three memory banks that hold channel \glspl{llr} for three polar code frames, as well as three additional memory banks to store the corresponding estimated codewords. The \gls{tcu} includes a test \gls{fsm} responsible to select the desired decoder, and to configure both the \gls{fll} and the decoders. In Fig.~\ref{fig:polarbear-arch}, the modules composing the \gls{tcu} have a dashed outline and are highlighted in blue.

The \gls{tcu} uses a serial interface to communicate with the outside world. This interface implements a simple protocol that allows to read and write to a memory map. As a consequence, we can communicate with the chip from a computer, e.g. to load the channel \glspl{llr} into the banks, to read back the content of the estimated codeword banks, and to configure the \gls{fll}.

\section{Test Chip and Measurement Results}\label{sec:results}
The \polarbear architecture described in Section~\ref{sec:chip_architecture} was fabricated in a 28\,nm FD-SOI CMOS technology, where the flexible decoder uses the regular $\text{V}_{\text{T}}$ flavor to minimize leakage and the unrolled decoder uses the low $\text{V}_{\text{T}}$ flavor to maximize speed. The other units present on the chip all use regular $\text{V}_{\text{T}}$. The core occupies 0.93\,mm$^2$ of the complete 1.47\,mm$^2$ die, and has an overall density of 62\%.

\begin{figure}[t]
	\centering
  \begin{tikzpicture}
    \node[anchor=south west, inner sep=0] (mg) at (0,0) {\includegraphics[width=\columnwidth]{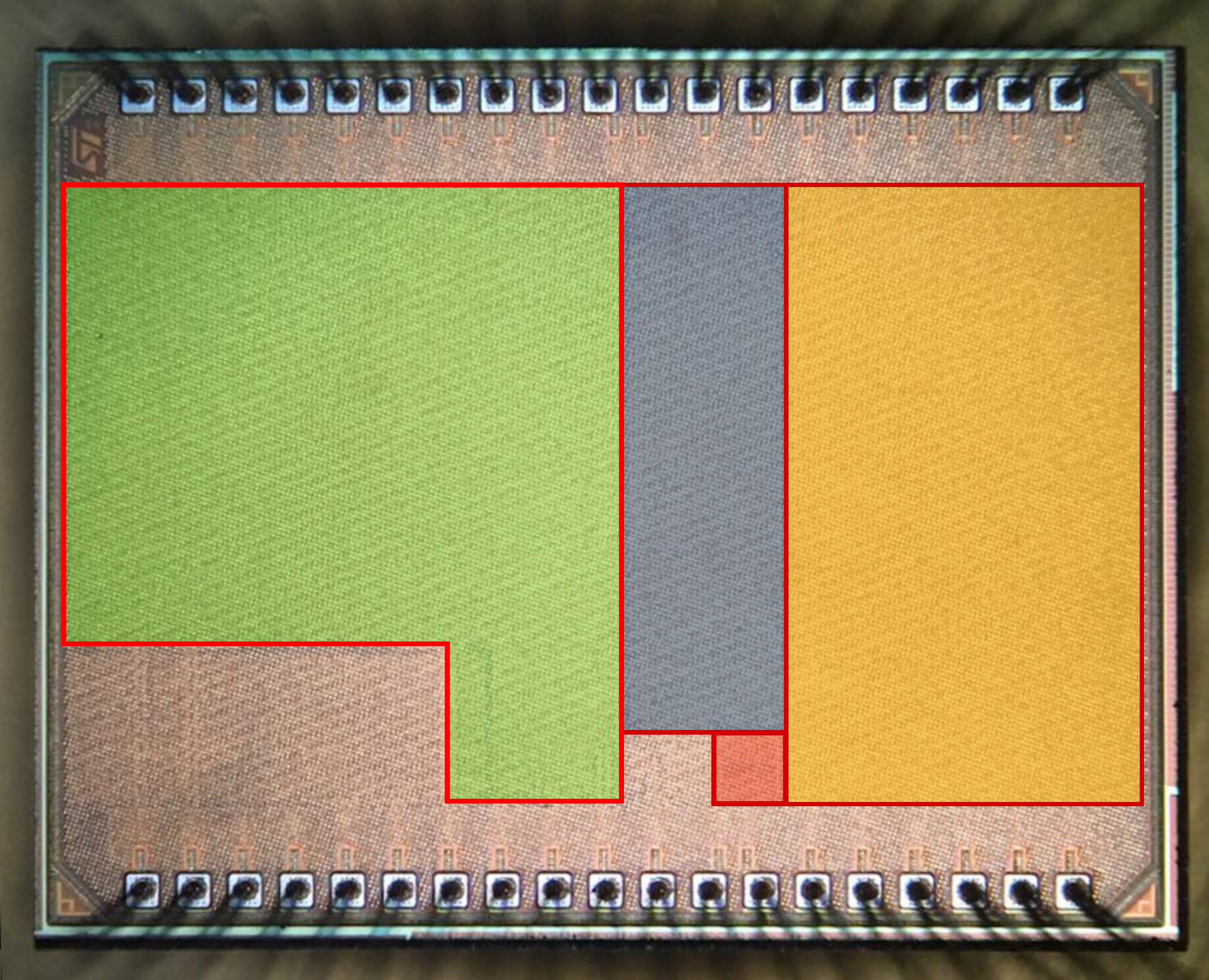}};
    \begin{scope}[x={(mg.south east)},y={(mg.north west)}]
      \node at (0.28,0.56) {\bf Flexible Decoder};
      \node at (0.28,0.51) {\footnotesize Reg. V$_\text{T}$, 0.44\,mm$^2$};
      \node at (0.8,0.58) {\bf Unrolled};
      \node at (0.8,0.53) {\bf Decoder};
      \node at (0.8,0.48) {\footnotesize Low V$_\text{T}$, 0.35\,mm$^2$};
      \node at (0.585,0.56) {\bf TCU};
      \node at (0.585,0.51) {\footnotesize 0.13\,mm$^2$};
      \draw[arrows=<-,line width=1.2pt] (0.62,0.21) -- (0.68, 0.25) node[right] {\hspace{-2pt}\bf CGU};
      \node at (0.74, 0.21) {\footnotesize 0.006\,mm$^2$};
      \coordinate (TL) at (0.04,0.965);
      \coordinate (TR) at (0.96,0.965);
      \draw[very thick,white] ($(TL)-(0,0.015)$) -- ($(TL)+(0,0.015)$) ($(TR)-(0,0.015)$) -- ($(TR)+(0,0.015)$);
      \path (TL) -- node[white] (diewidth) {\small $1400\,\mu$m} (TR);
      \draw[arrows=<-,line width=1.0pt,white] (TL) -- (diewidth);
      \draw[arrows=->,line width=1.0pt,white] (diewidth) -- (TR);
      \coordinate (TL2) at (0.02,0.935);
      \coordinate (BL) at (0.02,0.055);
      \draw[very thick,white] ($(TL2)-(0.015,0)$) -- ($(TL2)+(0.015,0)$) ($(BL)-(0.015,0)$) -- ($(BL)+(0.015,0)$);
      \path (TL2) -- node[white] (dieheight) {\rotatebox{90}{\small $1050\,\mu$m}} (BL);
      \draw[arrows=<-,line width=1.0pt,white] (TL2) -- (dieheight);
      \draw[arrows=->,line width=1.0pt,white] (dieheight) -- (BL);
      \end{scope}
  \end{tikzpicture}
	\caption{\polarbear micrograph.}
	\label{fig:chipoverview}\vspace{-3pt}
\end{figure}

Fig.~\ref{fig:chipoverview} shows a micrograph of the chip, where the area highlighted in green corresponds to the flexible decoder, the area in yellow is the fully-unrolled \gls{sc} decoder, the one in blue is the \gls{tcu} along with its memory, and the one in red is the \gls{cgu}. The \gls{cgu} can provide a clock frequency between 960\,kHz and 1.275\,GHz using an external reference clock of 20\,MHz and a supply voltage $\text{V}_{\text{CGU}}=0.9$\,V.

In the following sections, we start by describing our test setup and methodology. Then the various modes of the flexible decoder are compared against each other and against the unrolled SC decoder. Lastly our decoders are compared against the other fabricated polar decoders that can be found in the literature.

\subsection{Test Setup and Methodology}
Testing is conducted by inserting a \polarbear chip into a custom-made PCB which is, in turn, inserted as a daughterboard into an FPGA development board. The FPGA development board---a Xilinx XUPV5-LX110T---is connected to a PC via a serial interface. The steps to run a test can be summarized as follows:
\begin{enumerate}
\item Transfer the channel \glspl{llr} to the \gls{tcu} memory.
\item Configure the \gls{fll} to generate the desired fast clock.
\item Select the desired decoder (flexible or unrolled).
\item[] If the flexible decoder was selected:
  \begin{enumerate}
  \item Select the desired mode.
  \item Set the polar-code type: non-systematic or systematic.
  \item Select the \gls{crc} length-and-polynomial pair.
  \item Transfer the binary vector from which the set of frozen-bit indices $\mathcal{A}^c$ is derived.
  \item Set the index of the first information-bit location.
  \item Set the list size $L$ (SCL mode) or the maximum number of trials $T$ (SCF mode).
  \end{enumerate}
\item Start the test.
\item Wait until the decoder notifies the \gls{tcu} that decoding is complete.
\item Read the estimated codeword from the \gls{tcu} memory.
\item Compare the estimated codeword against the expected one.
\end{enumerate}

Measurement results are for test vectors generated using bit-true models of the decoders for an \gls{awgn} channel with an $\nicefrac{E_b}{N_0}$ of 0\,dB to obtain worst-case values, i.e., such that more switching activity is generated compared to operation in a typical $\nicefrac{E_b}{N_0}$ region of interest. Independent programmable power supplies are used to provide power to the various cores, and a high-precision multimeter is put in the loop to measure the current drawn by the decoder of interest. Furthermore, measurements are taken in continuous decoding mode at room temperature.

\begin{table}[t]
  \centering
  \caption{Decoding latency in clock cycles for the various supported decoders and modes corresponding to polar codes of 5 different code rates. The unrolled decoder is denoted U-SC.}
    \begin{tabular}{l|ccc}
      \toprule
      $R$               & SC   & SCL  & U-SC\\
      \hline
      $\nicefrac{1}{4}$ & 1577 & 1887 & -\\
      $\nicefrac{1}{2}$ & 1833 & 2408 & -\\
      $\nicefrac{2}{3}$ & 1896 & 2644 & -\\
      $\nicefrac{3}{4}$ & 1960 & 2783 & -\\
      $\nicefrac{5}{6}$ & 1991 & 2899 & 283\\
      \bottomrule
    \end{tabular}
  \label{tab:latency}
\end{table}

For reference, the latency---in clock cycles---of the polar codes used in the measurements are provided in Table~\ref{tab:latency}. The latency values for the \gls{scf} mode are not included in this table as they are integer multiples of those of the \gls{sc} decoder, where the multiplication factor is the number of trials. As it can be observed by combining equations~\eqref{eqn:latency:sc}, \eqref{eqn:latency:scl}, and \eqref{eqn:latency:scl:sort}, the latency and throughput of the \gls{scl} mode are independent of the list size $L$. This is a result of having all the necessary hardware resources to accommodate the largest supported list size $L_{\max}$.

From Table~\ref{tab:latency}, it can be seen that the latency increases with the code rate. The reason for that lies in the nature of \textit{good} polar codes where the first information bit location $b$ is pushed further and further to the right as the code rate $R$ decreases. As a result, the correction term of \eqref{eqn:latency:sc} increases as the code rate diminishes and the \gls{sc} latency $\mathcal{L}_{\text{sc}}$, common to all three modes, is reduced.

\subsection{Flexible Decoder}
The flexible decoder uses the regular V$_\text{T}$ process flavor, and occupies an area of 0.44\,mm$^2$ of which 0.29\,mm$^2$ are occupied by standard cells with a density of 65\%. The memory, in the form of registers, accounts for 26\% of the total flexible-decoder area.

\subsubsection{Quantization} 
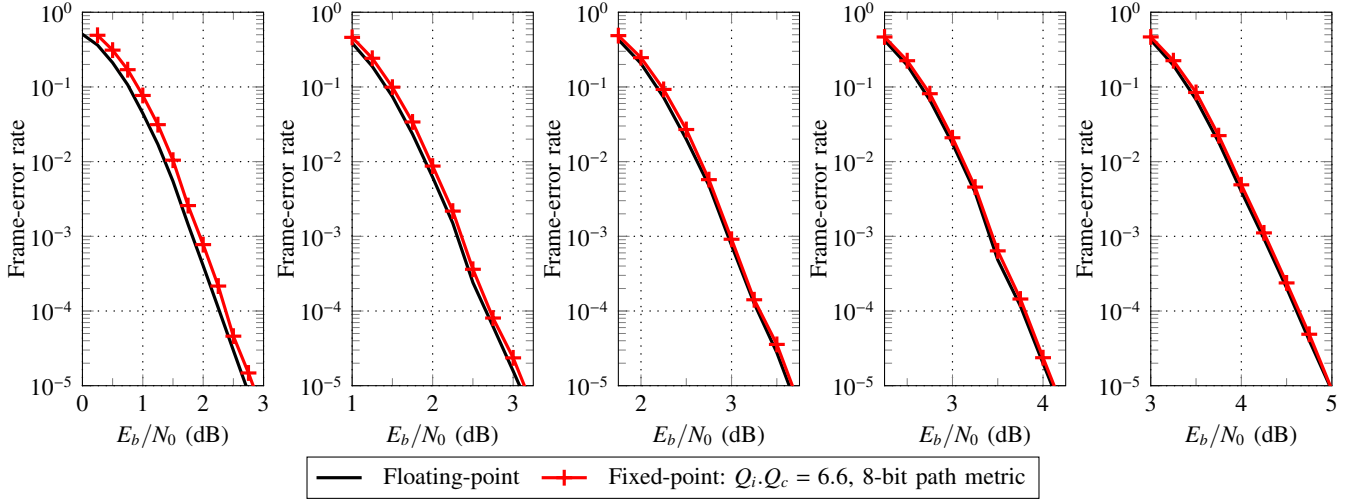
\begin{figure*}[t]
	\centering
  \begin{tikzpicture}

    \pgfplotsset{
      grid style = {
        dash pattern = on 0.05mm off 1mm,
        line cap = round,
        black,
        line width = 0.5pt
      },
      label style = {font=\fontsize{9pt}{7.2}\selectfont},
      tick label style = {font=\fontsize{9pt}{7.2}\selectfont}
    }

    \begin{semilogyaxis}[%
      xlabel=$E_b/N_0$ (dB),xtick={-1,0,1,2,3,...,6.0},%
      xlabel style={yshift=0.4em},%
      minor x tick num={1},
      xmin=0.0,xmax=3,%
      ymin=1e-5,ymax=1e0,%
      ylabel=Frame-error rate, ylabel style={yshift=-1.0em},%
      width=0.22\textwidth, height=6.55cm, grid=major,%
      legend style={
        anchor={center},
        cells={anchor=west},
        column sep=2mm,
        font=\fontsize{9pt}{7.2}\selectfont,
        mark size=3.0pt,
        mark options=solid
      },
      legend columns=2,
      legend to name=scl-qtz-legend,
      mark size=3.0pt,
      mark options=solid]
      
      \addplot[very thick,color=black] plot coordinates {
        (-5.00000e-01, 8.18404e-01)
        (-2.50000e-01, 6.76105e-01)
        (0.00000e+00, 5.07228e-01)
        (2.50000e-01, 3.63702e-01)
        (5.00000e-01, 2.11193e-01)
        (7.50000e-01, 1.06754e-01)
        (1.00000e+00, 4.43158e-02)
        (1.25000e+00, 1.70351e-02)
        (1.50000e+00, 5.45614e-03)
        (1.75000e+00, 1.44304e-03)
        (2.00000e+00, 4.04000e-04)
        (2.25000e+00, 1.10142e-04)
        (2.50000e+00, 2.96648e-05)
        (2.75000e+00, 8.09879e-06)
      };
      \addlegendentry{Floating-point}
      \addplot[very thick,color=red,mark=+] plot coordinates {
        (2.50000e-01, 4.91737e-01)
        (5.00000e-01, 3.11018e-01)
        (7.50000e-01, 1.70211e-01)
        (1.00000e+00, 7.68947e-02)
        (1.25000e+00, 3.12632e-02)
        (1.50000e+00, 1.04561e-02)
        (1.75000e+00, 2.57895e-03)
        (2.00000e+00, 7.74436e-04)
        (2.25000e+00, 2.15768e-04)
        (2.50000e+00, 4.60405e-05)
        (2.75000e+00, 1.48017e-05)
        (3.00000e+00, 4.19375e-06)
      };
      \addlegendentry{Fixed-point: $Q_i.Q_c=6.6$, 8-bit path metric}

    \end{semilogyaxis}
  \end{tikzpicture}\hspace{-7pt}
  \begin{tikzpicture}

    \pgfplotsset{
      grid style = {
        dash pattern = on 0.05mm off 1mm,
        line cap = round,
        black,
        line width = 0.5pt
      },
      label style = {font=\fontsize{9pt}{7.2}\selectfont},
      tick label style = {font=\fontsize{9pt}{7.2}\selectfont}
    }

    \begin{semilogyaxis}[%
      xlabel=$E_b/N_0$ (dB),xtick={0,1,2,3,...,6.0},%
      xlabel style={yshift=0.4em},%
      minor x tick num={1},
      xmin=1.0,xmax=3.25,%
      ymin=1e-5,ymax=1e-0,%
      ylabel=Frame-error rate, ylabel style={yshift=-1.0em},%
      width=0.22\textwidth, height=6.55cm, grid=major,%
      mark size=3.0pt,
      mark options=solid]
      
      \addplot[very thick,color=black] plot coordinates {
        (2.50000e-01, 9.22754e-01)
        (5.00000e-01, 7.90228e-01)
        (7.50000e-01, 5.83298e-01)
        (1.00000e+00, 3.86404e-01)
        (1.25000e+00, 1.88491e-01)
        (1.50000e+00, 7.43860e-02)
        (1.75000e+00, 2.33509e-02)
        (2.00000e+00, 6.17544e-03)
        (2.25000e+00, 1.50000e-03)
        (2.50000e+00, 2.40566e-04)
        (2.75000e+00, 6.20732e-05)
        (3.00000e+00, 1.57011e-05)
        (3.25000e+00, 3.84586e-06)
      };
      \addplot[very thick,color=red,mark=+] plot coordinates {
        (1.00000e+00, 4.61000e-01)
        (1.25000e+00, 2.42140e-01)
        (1.50000e+00, 9.92632e-02)
        (1.75000e+00, 3.39649e-02)
        (2.00000e+00, 8.71930e-03)
        (2.25000e+00, 2.17544e-03)
        (2.50000e+00, 3.62007e-04)
        (2.75000e+00, 8.05153e-05)
        (3.00000e+00, 2.35816e-05)
        (3.25000e+00, 5.04923e-06)
      };

    \end{semilogyaxis}
  \end{tikzpicture}\hspace{-3pt}
  \begin{tikzpicture}

    \pgfplotsset{
      grid style = {
        dash pattern = on 0.05mm off 1mm,
        line cap = round,
        black,
        line width = 0.5pt
      },
      label style = {font=\fontsize{9pt}{7.2}\selectfont},
      tick label style = {font=\fontsize{9pt}{7.2}\selectfont}
    }

    \begin{semilogyaxis}[%
      xlabel=$E_b/N_0$ (dB),xtick={0,1,2,3,...,6.0},%
      xlabel style={yshift=0.4em},%
      minor x tick num={1},
      xmin=1.75,xmax=3.75,%
      ymin=1e-5,ymax=1e-0,%
      ylabel=Frame-error rate, ylabel style={yshift=-1.0em},%
      width=0.22\textwidth, height=6.55cm, grid=major,%
      mark size=3.0pt,
      mark options=solid]
      
      \addplot[very thick,color=black] plot coordinates {
        (1.00000e+00, 9.64333e-01)
        (1.25000e+00, 8.72053e-01)
        (1.50000e+00, 6.84930e-01)
        (1.75000e+00, 4.30526e-01)
        (2.00000e+00, 2.05702e-01)
        (2.25000e+00, 7.23509e-02)
        (2.50000e+00, 1.99298e-02)
        (2.75000e+00, 4.78947e-03)
        (3.00000e+00, 7.55396e-04)
        (3.25000e+00, 1.26092e-04)
        (3.50000e+00, 2.76855e-05)
        (3.75000e+00, 4.31648e-06)
      };
      \addplot[very thick,color=red,mark=+] plot coordinates {
        (7.50000e-01, 9.95947e-01)
        (1.00000e+00, 9.76333e-01)
        (1.25000e+00, 9.03123e-01)
        (1.50000e+00, 7.37386e-01)
        (1.75000e+00, 4.86316e-01)
        (2.00000e+00, 2.46526e-01)
        (2.25000e+00, 9.26316e-02)
        (2.50000e+00, 2.69123e-02)
        (2.75000e+00, 5.73684e-03)
        (3.00000e+00, 9.13793e-04)
        (3.25000e+00, 1.41457e-04)
        (3.50000e+00, 3.57017e-05)
        (3.75000e+00, 5.56321e-06)
      };

    \end{semilogyaxis}
  \end{tikzpicture}\hspace{-3pt}
  \begin{tikzpicture}

    \pgfplotsset{
      grid style = {
        dash pattern = on 0.05mm off 1mm,
        line cap = round,
        black,
        line width = 0.5pt
      },
      label style = {font=\fontsize{9pt}{7.2}\selectfont},
      tick label style = {font=\fontsize{9pt}{7.2}\selectfont}
    }

    \begin{semilogyaxis}[%
      xlabel=$E_b/N_0$ (dB),xtick={0,1,2,3,...,6.0},%
      xlabel style={yshift=0.4em},%
      minor x tick num={1},
      xmin=2.25,xmax=4.25,%
      ymin=1e-5,ymax=1e-0,%
      ylabel=Frame-error rate, ylabel style={yshift=-1.0em},%
      width=0.22\textwidth, height=6.55cm, grid=major,%
      mark size=3.0pt,
      mark options=solid]
      
      \addplot[very thick,color=black] plot coordinates {
        (1.50000e+00, 9.66965e-01)
        (1.75000e+00, 8.76281e-01)
        (2.00000e+00, 6.82281e-01)
        (2.25000e+00, 4.20912e-01)
        (2.50000e+00, 1.95702e-01)
        (2.75000e+00, 6.66667e-02)
        (3.00000e+00, 1.74737e-02)
        (3.25000e+00, 3.87719e-03)
        (3.50000e+00, 4.83871e-04)
        (3.75000e+00, 1.16494e-04)
        (4.00000e+00, 1.99045e-05)
        (4.25000e+00, 3.51605e-06)
      };
      \addplot[very thick,color=red,mark=+] plot coordinates {
        (1.50000e+00, 9.76140e-01)
        (1.75000e+00, 9.00807e-01)
        (2.00000e+00, 7.21947e-01)
        (2.25000e+00, 4.66596e-01)
        (2.50000e+00, 2.24491e-01)
        (2.75000e+00, 8.11754e-02)
        (3.00000e+00, 2.08772e-02)
        (3.25000e+00, 4.56140e-03)
        (3.50000e+00, 6.38554e-04)
        (3.75000e+00, 1.44718e-04)
        (4.00000e+00, 2.36967e-05)
        (4.25000e+00, 4.06240e-06)
      };

    \end{semilogyaxis}
  \end{tikzpicture}\hspace{-3pt}
  \begin{tikzpicture}

    \pgfplotsset{
      grid style = {
        dash pattern = on 0.05mm off 1mm,
        line cap = round,
        black,
        line width = 0.5pt
      },
      label style = {font=\fontsize{9pt}{7.2}\selectfont},
      tick label style = {font=\fontsize{9pt}{7.2}\selectfont}
    }

    \begin{semilogyaxis}[%
      xlabel=$E_b/N_0$ (dB),xtick={0,1,2,3,...,6.0},%
      xlabel style={yshift=0.4em},%
      minor x tick num={1},
      xmin=3.0,xmax=5,%
      ymin=1e-5,ymax=1e-0,%
      ylabel=Frame-error rate, ylabel style={yshift=-1.0em},%
      width=0.22\textwidth, height=6.55cm, grid=major,%
      mark size=3.0pt,
      mark options=solid]
      
      \addplot[very thick,color=black] plot coordinates {
        (2.25000e+00, 9.67941e-01)
        (2.50000e+00, 8.77000e-01)
        (2.75000e+00, 6.80882e-01)
        (3.00000e+00, 4.25235e-01)
        (3.25000e+00, 1.94706e-01)
        (3.50000e+00, 6.78824e-02)
        (3.75000e+00, 1.82353e-02)
        (4.00000e+00, 4.00000e-03)
        (4.25000e+00, 9.29204e-04)
        (4.50000e+00, 2.03593e-04)
        (4.75000e+00, 4.03551e-05)
        (5.00000e+00, 8.59476e-06)
      };
      \addplot[very thick,color=red,mark=+] plot coordinates {
        (2.00000e+00, 9.96294e-01)
        (2.25000e+00, 9.76000e-01)
        (2.50000e+00, 9.00000e-01)
        (2.75000e+00, 7.23588e-01)
        (3.00000e+00, 4.67176e-01)
        (3.25000e+00, 2.24647e-01)
        (3.50000e+00, 8.43529e-02)
        (3.75000e+00, 2.22941e-02)
        (4.00000e+00, 4.89286e-03)
        (4.25000e+00, 1.11111e-03)
        (4.50000e+00, 2.37647e-04)
        (4.75000e+00, 4.88395e-05)
        (5.00000e+00, 8.82768e-06)
      };

    \end{semilogyaxis}
  \end{tikzpicture}
  \\
  \ref{scl-qtz-legend}
	\caption{Impact of LLR and path metric quantization on the error-correction performance of 8-bit CRC-aided SCL decoding with $L=4$. From left to right, the performance of polar codes of blocklength $N=1024$ with various code rates $R \in \{ \nicefrac{1}{4}, \nicefrac{1}{2}, \nicefrac{2}{3}, \nicefrac{3}{4}, \nicefrac{5}{6}\}$.}
	\label{fig:scl-qtz}\vspace{-3pt}
\end{figure*}

In terms of quantization, this decoder uses $Q_i$.$Q_c$ equal to 6.6, and 8-bit path metrics for the \gls{scl} mode. Fig.~\ref{fig:scl-qtz} shows the impact of this quantization on the error-correction performance of 8-bit CRC-aided SCL decoding with $L=4$ for polar codes of various rates. It can be seen that this quantization incurs a coding loss ranging from 0.13\,dB to under 0.05\,dB, at a \gls{fer} of $10^{-3}$, compared to using a floating-point representation. We note that the coding loss is greater for the lower-rate codes and diminishes as the rate increases.

\begin{table}[t]
  \centering
  \caption{CRC lengths and polynomials supported by the flexible decoder.}
    \begin{tabular}{cc}
      \toprule
      \textbf{Length}& \multirow{2}{*}{\textbf{Polynomial}}\\
      (bits)\\
      \hline
      4  &  $x^4 + x + 1$ \\
      8  &  $x^8 + x^7 + x^4 + x^2 + x + 1$ \\
      16 &  $x^{16} + x^{12} + x^5 + 1$ \\
      \bottomrule
    \end{tabular}
  \label{tab:crcpoly}
\end{table}

\subsubsection{Decoding Modes} As mentioned earlier, the flexible decoder has three operating modes corresponding to the \gls{sc}, \gls{scf}, and \gls{scl} algorithms. The operating mode can be selected at execution time.

The \gls{scl} mode supports a list size $L$ value up to $L_{\max}=4$. As can be seen from Fig.~\ref{fig:ec-perf-cmp}, for an $N = 1024$ polar code, moving from $L=4$ to $L=8$ (or even $L=32$) results in a small gain in terms of the error-correction performance for this particular code rate and we observe similar behavior for other code rates. This fact, combined with the area constraints we had for our chip, lead to the choice of $L_{\max}=4$. Since in our architecture the configured list size $L$ has to be a power of two, our chip supports the list sizes $L \in \{1,2,4\}$, where $L=1$ is equivalent to \gls{sc} mode selection. The \gls{crc} lengths supported by the decoder chip, which can be selected at the time of execution, are summarized in Table~\ref{tab:crcpoly} along with the \gls{crc} polynomials that were used. These lengths were selected to cover a wide range of list sizes and rates, as different operating conditions require different \gls{crc} lengths in order to achieve the best possible performance~\cite{Balatsoukas_TSP_2015}. We note that, for \gls{scl} decoding it is also possible to completely disable the \gls{crc}.

In the \gls{scf} mode, the maximum number of trials $T$ has to be set and can have a value of up to $T_{\max}$. As can be seen from Fig.~\ref{fig:ec-perf-cmp}, for an $N = 1024$ polar code, moving from $T=8$ to $T=16$ provides very little benefit in terms of the error-correction performance. However, since increasing $T_{\max}$ incurs a negigible hardware overhead because the LLR sorter area is very small, we decided to choose $T_{\max}=32$ in order to ensure that we can cover a very wide range of code rate scenarios. While it is optional in the \gls{scl} mode, the \gls{scf} mode mandates activation of a \gls{crc} unit and the selection of a \gls{crc} length.

The \gls{sc} mode can be selected by disabling the \gls{crc} and setting $L = 1$.

Regarding the critical path of the flexible decoder, it depends on the operating mode and parameters. In \gls{scl} mode with a list size $L=4$, the critical path starts at the output of a register storing a path metric, goes through the metric sorter, then through a \gls{psn} (part of a decoder core), and ends at the input of the path-memory register. For the \gls{sc} and \gls{scf} modes as well as the \gls{scl} mode with $L=2$, the critical path starts from an internal-\gls{llr} memory register, goes through a processing element and into the \gls{psn} (both part of a decoder core) and ends at the input of a path-memory register.

As for any polar decoder, the flexible decoder can decode polar codes with blocklengths $N$ smaller than 1024 by setting the $1024-N$ most significant channel-\gls{llr} locations to the fixed-point equivalent of $+\infty$. However, since the controller was not optimized towards this goal, minute changes to its architecture would be required to achieve the optimal latency with no noticeable impact on area or clock frequency.

\subsubsection{Throughput Comparison}\label{sec:results:tp}

In this section, the measured throughput and energy per bit of the three modes are compared. The 8-bit \gls{crc} is selected for the \gls{scf} and \gls{scl} modes. Since the throughput, and thus the energy per bit, of the \gls{scf} mode are highly dependent on the average number of trials, results are provided for the average number of trials required at two \gls{fer} values of interest.

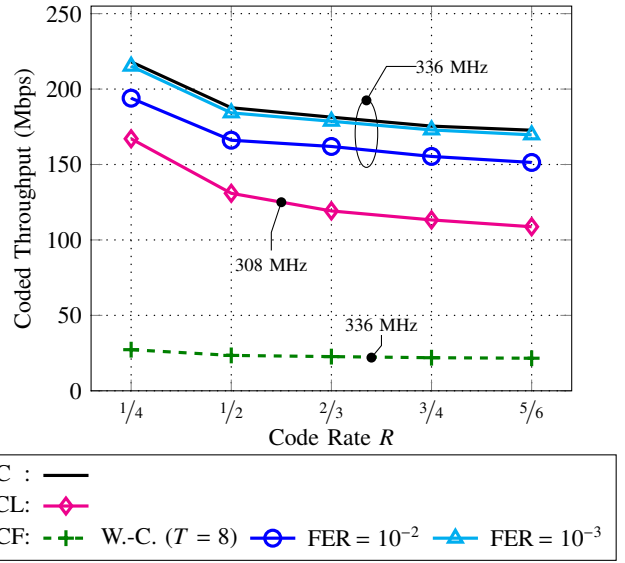
\begin{figure}[t]
	\centering
  \begin{tikzpicture}

    \pgfplotsset{
      grid style = {
        dash pattern = on 0.05mm off 1mm,
        line cap = round,
        black,
        line width = 0.5pt
      },
      label style = {font=\fontsize{9pt}{7.2}\selectfont},
      tick label style = {font=\fontsize{9pt}{7.2}\selectfont}
    }

    \tikzset{
      branch/.style={fill,shape=circle,scale=0.4},
    }

    \begin{axis}[%
      xlabel={Code Rate $R$},xtick=data,%
      xticklabels={{$\nicefrac{1}{4}$},{$\nicefrac{1}{2}$},{$\nicefrac{2}{3}$},{$\nicefrac{3}{4}$},{$\nicefrac{5}{6}$}},%
      xlabel style={yshift=0.4em},%
      ymin=0,ymax=255,%
      ylabel={Coded Throughput (Mbps)}, ylabel style={yshift=-1.0em},%
      width=0.9\columnwidth, height=6.7cm, grid=major,%
      legend style={
        anchor={center},
        cells={anchor=west},
        column sep=0.75mm,
        font=\fontsize{9pt}{7.2}\selectfont,
        mark size=3.0pt,
        mark options=solid
      },
      legend columns=4,
      legend to name=tp-legend,
      mark size=3.0pt,
      mark options=solid
      ]
      
      \addlegendimage{empty legend}
      \addlegendentry[anchor=east]{SC\phantom{F}:}

      \addplot[very thick,black] plot coordinates {
        (1,218.06)
        (2,187.61)
        (3,181.37)
        (4,175.45)
        (5,172.72)
      };
      \addlegendentry{}
      \addlegendimage{empty legend}
      \addlegendentry{}
      \addlegendimage{empty legend}
      \addlegendentry{}

      \draw (axis cs:3.35,170.5) ellipse (0.15cm and 0.45cm);
      \draw[-] (axis cs:3.35,192.5) node[branch] {} -- (axis cs:3.55,215) -- (axis cs:3.75,215) node[right] {\scriptsize 336~MHz} -- (axis cs:3.85,215) ;
      
      \addlegendimage{empty legend}
      \addlegendentry[anchor=east]{SCL:}
      \addplot[very thick,color=magenta, mark=diamond] plot coordinates {
        (1,167.05)
        (2,130.91)
        (3,119.22)
        (4,113.27)
        (5,108.74)
      };
      \addlegendentry{}
      \addlegendimage{empty legend}
      \addlegendentry{}
      \addlegendimage{empty legend}
      \addlegendentry{}

      \draw[-] (axis cs:2.5,125) node[branch] {} -- (axis cs:2.4,105) -- (axis cs:2.4,94) node[below] {\scriptsize 308~MHz} -- (axis cs:2.4,93);

      \addlegendimage{empty legend}
      \addlegendentry[anchor=east]{SCF:}
      \addplot[very thick, dashed, color=darkgreen, mark=+] plot coordinates {
        (1,27.26)
        (2,23.45)
        (3,22.67)
        (4,21.93)
        (5,21.59)
      };
      \addlegendentry{W.-C. ($T=8$)}

      \addplot[very thick,color=blue, mark=o] plot coordinates {
        (1,193.97)
        (2,166.10)
        (3,161.95)
        (4,155.35)
        (5,151.36)
      };
      \addlegendentry{FER$\,=10^{-2}$}

      \addplot[very thick,color=cyan, mark=triangle] plot coordinates {
        (1,215.16)
        (2,184.23)
        (3,178.67)
        (4,172.92)
        (5,169.65)
      };
      \addlegendentry{FER$\,=10^{-3}$}

      \draw[-] (axis cs:3.5,54) node[below] {\scriptsize 336~MHz};
      \draw[-] (axis cs:3.40,22) node[branch] {} -- (axis cs:3.5,30) -- (axis cs:3.5,37);

    \end{axis}

  \end{tikzpicture}
  \\
  \ref{tp-legend}
	\caption{Coded throughput to decode polar codes of blocklength $N=1024$ using all three modes supported by the flexible decoder. Maximum achievable clock frequencies $f_{\text{clk}}$ shown as annotations.}
	\label{fig:tp-cmp}\vspace{-3pt}
\end{figure}

Fig.~\ref{fig:tp-cmp} shows the throughput for the three modes supported by the flexible decoder.
All measurements are for the same core supply voltage of 0.9\,V and for the respective maximum achievable clock frequency. 
Fig.~\ref{fig:tp-cmp} shows that the \gls{sc} mode has a throughput that is from 31\% to 59\% greater than that of the \gls{scl} mode. While the \gls{wc} throughput of the \gls{scf} mode is well below that of any other mode, the achievable throughput of the \gls{scf} mode approaches that of the \gls{sc} mode as the \gls{fer} improves. While operating at a \gls{fer} of $10^{-2}$, the \gls{scf} mode is approximately 12\% slower than the \gls{sc} mode. This gap shrinks to under 1.5\% at a \gls{fer} of $10^{-3}$. Comparing the \gls{scf} mode at a \gls{fer} of $10^{-2}$ with the \gls{scl} mode, the \gls{scf} mode is from 16\% to 39\% faster than the \gls{scl} mode for the lowest to the highest code rates, respectively.

\subsubsection{Energy-per-bit Comparison}\label{sec:results:energy}
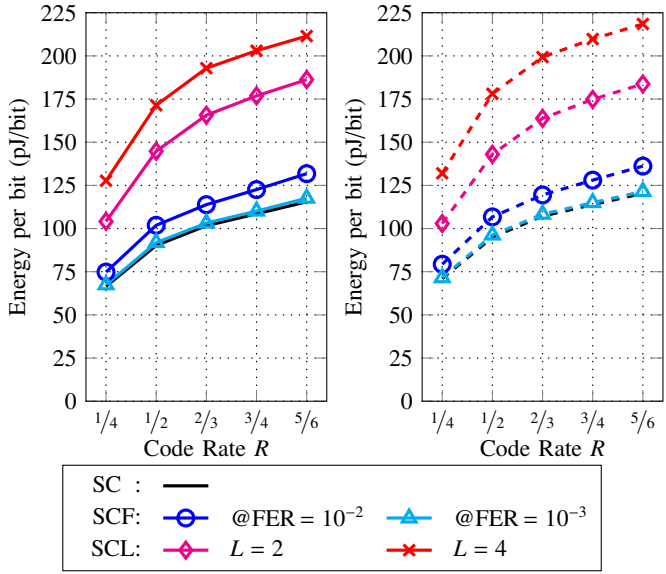
\begin{figure}[t]
	\centering
  \begin{tikzpicture}

    \pgfplotsset{
      grid style = {
        dash pattern = on 0.05mm off 1mm,
        line cap = round,
        black,
        line width = 0.5pt
      },
      label style = {font=\fontsize{9pt}{7.2}\selectfont},
      tick label style = {font=\fontsize{9pt}{7.2}\selectfont}
    }

    \begin{axis}[%
      xlabel={Code Rate $R$},xtick=data,%
      xticklabels={{$\nicefrac{1}{4}$},{$\nicefrac{1}{2}$},{$\nicefrac{2}{3}$},{$\nicefrac{3}{4}$},{$\nicefrac{5}{6}$}},%
      xlabel style={yshift=0.4em},%
      ymin=0,ymax=225,%
      ytick={0,25,50,...,225},
      ylabel={Energy per bit (pJ/bit)}, ylabel style={yshift=-1.0em},%
      width=0.54\columnwidth, height=6.75cm, grid=major,%
      legend style={
        anchor={center},
        cells={anchor=west},
        column sep=2mm,
        font=\fontsize{9pt}{7.2}\selectfont,
        mark size=3.0pt
      },
      legend columns=3,
      legend to name=energy-eff-legend,
      mark size=3.0pt]
      
      \addlegendimage{empty legend}
      \addlegendentry[anchor=east]{SC\phantom{F}:}

      \addplot[very thick,black] plot coordinates {
        (1, 66.53)
        (2, 90.22)
        (3,101.65)
        (4,108.53)
        (5,115.49)
      };
      \addlegendentry{}

      \addlegendimage{empty legend}
      \addlegendentry{}

      \addlegendimage{empty legend}
      \addlegendentry[anchor=east]{SCF:}

      \addplot[very thick,color=blue, mark=o] plot coordinates {
        (1, 74.79)
        (2,101.90)
        (3,113.84)
        (4,122.57)
        (5,131.79)
      };
      \addlegendentry{@FER$\,=10^{-2}$}

      \addplot[very thick,color=cyan, mark=triangle] plot coordinates {
        (1, 67.43)
        (2, 91.87)
        (3,103.19)
        (4,110.11)
        (5,117.58)
      };
      \addlegendentry{@FER$\,=10^{-3}$}

      \addlegendimage{empty legend}
      \addlegendentry[anchor=east]{SCL:}

      \addplot[very thick,color=magenta, mark=diamond] plot coordinates {
        (1,104.15)
        (2,144.76)
        (3,165.69)
        (4,176.85)
        (5,186.26)
      };
      \addlegendentry{$L = 2$}

      \addplot[very thick,color=red, mark=x] plot coordinates {
        (1,127.70)
        (2,171.43)
        (3,192.88)
        (4,203.02)
        (5,211.48)
      };
      \addlegendentry{$L = 4$}

    \end{axis}
  \end{tikzpicture}
  \begin{tikzpicture}

    \pgfplotsset{
      grid style = {
        dash pattern = on 0.05mm off 1mm,
        line cap = round,
        black,
        line width = 0.5pt
      },
      label style = {font=\fontsize{9pt}{7.2}\selectfont},
      tick label style = {font=\fontsize{9pt}{7.2}\selectfont}
    }

    \begin{axis}[%
      xlabel={Code Rate $R$},xtick=data,%
      xticklabels={{$\nicefrac{1}{4}$},{$\nicefrac{1}{2}$},{$\nicefrac{2}{3}$},{$\nicefrac{3}{4}$},{$\nicefrac{5}{6}$}},%
      xlabel style={yshift=0.4em},%
      ymin=0,ymax=225,%
      ytick={0,25,50,...,225},
      ylabel={Energy per bit (pJ/bit)}, ylabel style={yshift=-1.0em},%
      width=0.54\columnwidth, height=6.75cm, grid=major,%
      mark size=3.0pt,
      mark options=solid]

      \addplot[very thick,dashed,black] plot coordinates {
        (1, 70.99)
        (2, 94.99)
        (3,107.18)
        (4,113.88)
        (5,120.37)
      };

      \addplot[very thick,dashed,color=blue, mark=o] plot coordinates {
        (1, 79.34)
        (2,106.75)
        (3,119.48)
        (4,128.04)
        (5,136.16)
      };

      \addplot[very thick,dashed,color=cyan, mark=triangle] plot coordinates {
        (1, 71.53)
        (2, 96.24)
        (3,108.30)
        (4,115.02)
        (5,121.49)
      };

      \addplot[very thick,dashed,color=magenta, mark=diamond] plot coordinates {
        (1,102.90)
        (2,143.00)
        (3,163.81)
        (4,174.81)
        (5,183.75)
      };

      \addplot[very thick,dashed,color=red, mark=x] plot coordinates {
        (1,132.00)
        (2,178.07)
        (3,199.29)
        (4,209.77)
        (5,218.51)
      };
      
    \end{axis}
  \end{tikzpicture}
  \\
  \ref{energy-eff-legend}
	\caption{Energy per bit to decode polar codes of blocklength $N=1024$ using the various decoding algorithms supported by the flexible decoder. All measurements are for a core supply voltage of 0.9\,V. Results on the left (solid curves) are for a clock $f_{\text{clk}}=100$\,MHz while the ones on the right (dashed curves) are for the respective maximum achievable clock frequency, i.e., $f_{\text{clk}}=336$\,MHz for both SC and SCF modes, and 308\,MHz for the SCL mode.}
	\label{fig:energy-eff-cmp}\vspace{-3pt}
\end{figure}

Fig.~\ref{fig:energy-eff-cmp} shows the energy efficiency for the various modes supported by the flexible decoder. For fair comparison, all measurements are for the same core supply voltage of 0.9\,V. The solid curves on the left-hand side of the figure are all for a clock frequency of $f_{\text{clk}}=100$\,MHz whereas the dashed curves on the right-hand side of the figure are for the maximum achievable clock frequencies for each decoder and mode. An 8-bit \gls{crc} is used for the \gls{scf} and \gls{scl} decoders. The energy per bit is defined as:
\begin{equation}
  \frac{\text{Power } (W)}{\text{Coded T/P } (bps)}.\nonumber
\end{equation}

From both sides of Fig.~\ref{fig:energy-eff-cmp} we observe that more energy is required as the code rate increases regardless of the operating mode. This is an expected result as the latency (number of required \glspl{cc}) increases with the code rate, as can be seen from Table~\ref{tab:latency}. The \gls{scl} mode has the greatest latency among the three modes and uses the majority of the modules of the flexible decoder illustrated in Fig.~\ref{fig:flexible-arch}. Thus, as expected, Fig.~\ref{fig:energy-eff-cmp} shows that, indeed, the \gls{scl} mode requires the most energy out of the three supported modes.
From the same figure, we observe that the energy per bit of the \gls{scf} mode approaches that of the \gls{sc} decoder as the \gls{fer} improves (or as the $\nicefrac{E_b}{N_0}$ ratio increases). 

\subsubsection{Discussion}
With three modes that offer different characteristics, the adequate configuration can be selected at execution time according to the requirements and operating conditions.
The \gls{sc} mode has a constant latency, and the best throughput and energy per bit.
The \gls{scl} mode, with a list size $L=4$, requires from $1.8\times$ to $1.9\times$ more energy per bit as the \gls{sc} mode, but its error-correction performance is significantly better than that of \gls{sc}. With an error-correction performance that approaches that of the \gls{scl} algorithm with $L=2$ and an average throughput that tends to that of the \gls{sc} mode as the signal-to-noise ratio improves, the \gls{scf} mode appears as the most attractive mode if the decoder is operated in a good $\nicefrac{E_b}{N_0}$ region and if the system can cope with the variable execution time.
It is interesting to note that \gls{scl} decoding with $L=4$ does not require twice as much energy per bit than with $L=2$. The energy-per-bit gap between the \gls{sc} mode and the \gls{scl} mode with $L=2$ is greater. The initial energy hit comes from the greater latency of \gls{scl} decoding combine with the increase in hardware resources used. Increasing $L$ from 2 to 4, the latency remains unchanged, only the additional hardware resources used contribute to increase the energy required per bit.

\begin{table*}[t]
  \centering
  \caption{Comparison of the flexible decoder against the other fabricated ASIC decoders for a (1024,\,512) polar code. An 8-bit \gls{crc} is used for the \gls{scf} and \gls{scl} decoders.}
  \begin{tabular}{l|c|c|ccc|c|cc}
    \toprule
    \hline
    \textbf{Implementation}& \multicolumn{5}{c|}{\textbf{This work}} & \cite{Mishra2012}  & \multicolumn{2}{c}{\cite{Park2014}}\\
    \hline\hline
    \textbf{Algorithm}          & SC& SCF ($T=8$) & \multicolumn{3}{c|}{SCL ($L=4$)} & SC   & \multicolumn{2}{c}{BP (15 iter.)}\\\hline
    \textbf{FER @ $\nicefrac{E_b}{N_0}=4$\,dB}
                           &$\sim 1 \times 10^{-5}$& $\sim 1 \times 10^{-7}$&\multicolumn{3}{c|}{$\sim 7 \times 10^{-8}$}&$\sim 1 \times 10^{-5}$& \multicolumn{2}{c}{$\sim 7 \times 10^{-5}$}\\\hline
	  $\nicefrac{E_b}{N_0}$ @ \textbf{FER of $= 10^{-5}$}
                           &$\sim 4$\,dB & $\sim 3.4$\,dB & \multicolumn{3}{c|}{$\sim 3$\,dB} & $\sim 4$\,dB & \multicolumn{2}{c}{$\sim 4.8$\,dB}\\\hline
    \textbf{Technology}          &28\,nm & 28\,nm & \multicolumn{3}{c|}{28\,nm}& 180\,nm   & \multicolumn{2}{c}{65\,nm}\\\hline
    \textbf{Area} (mm$^2$)       & 0.44$^a$ & 0.44$^a$ & \multicolumn{3}{c|}{0.44$^a$} & 1.71     & \multicolumn{2}{c}{1.48}\\\hline
    \textbf{Supply} (V)          & 0.9   & 0.9 & 1.3 & 0.9 & 0.5 & 1.3      & 1.0 & 0.475\\\hline
    \textbf{Frequency} (MHz)     &336&336&721&308&20& 150      & 300 & 50\\\hline
    \multirow{2}{*}{\textbf{Latency} }\multirow{2}{*}{\Shortunderstack{(CCs)\\($\mu$s)}}%
                           & 1\,833 &14\,664 (1\,833$^b$)&\multicolumn{3}{c|}{2\,408}&1\,568& \multicolumn{2}{c}{150 (65.7$^b$)}\\\cline{2-9}
                           &5.46&43.67&3.34&7.82&120.40& 10.45    & 0.50 & 3.00\\\hline
    \textbf{Coded T/P} (Mbps)    &187.6& \phantom{$^b$}187.6$^b$&306.8&130.9&8.5& 98.0  & \phantom{$^{b,c}$}4\,675.8$^{b,c}$ & \phantom{$^{b,c}$}779.3$^{b,c}$ \\\hline
    \textbf{W.-C. Coded T/P} (Mbps)
                           &187.6&23.5&306.8&130.9&8.5& 98.0  & 2\,048.0 & 341.3 \\\hline
    \textbf{Area Eff.} (Mbps/mm$^2$)&423& \phantom{$^b$}423$^b$&692&295&19& 57 & \phantom{$^{b,c}$}3,168$^{b,c}$ & \phantom{$^{b,c}$}528$^{b,c}$\\\hline
    \textbf{Power} (mW)          &17.8&17.9&128.3&23.3&0.6& 67    & 477.5 & 18.6 \\\hline
    \textbf{Energy per bit} (pJ/bit)&95.0& \phantom{$^b$}95.5$^b$&418.3&178.1&64.7& 684 & \phantom{$^{b,c}$}102.1$^{b,c}$ & \phantom{$^{b,c}$}23.8$^{b,c}$ \\\hline
    \hline
    \multicolumn{8}{l}{\textit{Normalized for $28$\,nm and $0.9$\,V$\,^\diamond$}}\\
    \hline
    \hline
    \textbf{Area} (mm$^2$)       &0.44$^a$&0.44$^a$&\multicolumn{3}{c|}{0.44$^a$}&0.04&\multicolumn{2}{c}{0.27}\\\hline
    \textbf{Frequency} (MHz)     &336&336&--&308&--&1\,335&696 &--\\\hline
    \textbf{Latency} ($\mu$s)    &5.46&43.67&--&7.82&--&1.17&0.22&--\\\hline
    \textbf{Coded T/P} (Mbps)    &187.6& \phantom{$^b$}187.6$^b$&--&130.9&--&871.9& \phantom{$^{b,c}$}10\,847.9$^{b,c}$&--\\\hline
    \textbf{W.-C. Coded T/P} (Mbps)
                           &187.6&23.5&--&130.9&--&871.9&4\,751.4&--\\\hline
    \textbf{Area Eff.} (Mbps/mm$^2$)&423& \phantom{$^b$}423$^b$&--&295&--& 21\,073 & \phantom{$^{b,c}$}39\,500$^{b,c}$& --\\\hline
    \textbf{Power} (mW)          &17.8&17.9&--&23.3&--& 5.0 & 166.6 & -- \\\hline
    \textbf{Energy per bit} (pJ/bit)&95.0& \phantom{$^b$}95.5$^b$&--&178.1&--& 5.7 & \phantom{$^{b,c}$}15.4$^{b,c}$& --\\\hline
    \bottomrule
    \multicolumn{9}{l}{$^a$All three modes supported by our flexible decoder occupy the same 0.44\,mm$^2$.}\\
    \multicolumn{9}{l}{$^b$Average value at $\nicefrac{E_b}{N_0}=4$\,dB.}\\
    \multicolumn{9}{l}{$^c$With early-termination and an average number of iterations of 6.57.}\\
    \multicolumn{9}{l}{$^\diamond$Area scaled as $s^2$, frequency as $\nicefrac{1}{s}$, and power as $v^2s$, where $s$ is the technology feature size and $v$ is the supply voltage ratio.}\\
    \multicolumn{9}{l}{\phantom{$^\diamond$}The frequency of \cite{Mishra2012} was first scaled back linearly to $1.8$\,V, the nominal voltage of the 180\,nm technology.}\\
  \end{tabular}\vspace{-11pt}
  \label{tab:asic:flexible}
\end{table*}

\subsection{Fully-Unrolled Partially-Pipelined SC Decoder}

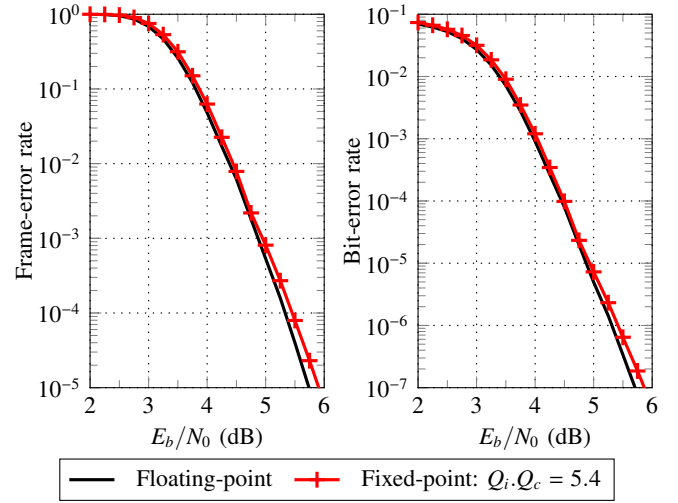
\begin{figure}[t]
	\centering
  \begin{tikzpicture}

    \pgfplotsset{
      grid style = {
        dash pattern = on 0.05mm off 1mm,
        line cap = round,
        black,
        line width = 0.5pt
      },
      label style = {font=\fontsize{9pt}{7.2}\selectfont},
      tick label style = {font=\fontsize{9pt}{7.2}\selectfont}
    }

    \begin{semilogyaxis}[%
      xlabel=$E_b/N_0$ (dB),xtick={2,3,...,6.0},%
      xlabel style={yshift=0.4em},%
      minor x tick num={1},
      xmin=2,xmax=6,%
      ymin=1e-5,ymax=1e0,%
      ylabel=Frame-error rate, ylabel style={yshift=-1.0em},%
      width=0.53\columnwidth, height=6.55cm, grid=major,%
      legend style={
        anchor={center},
        cells={anchor=west},
        column sep=2mm,
        font=\fontsize{9pt}{7.2}\selectfont,
        mark size=3.0pt,
        mark options=solid
      },
      legend columns=2,
      legend to name=unrolled-perf-legend,
      mark size=3.0pt,
      mark options=solid]
      
      \addplot[very thick,color=black] plot coordinates {
        (2.00000e+00, 9.98879e-01)
        (2.25000e+00, 9.90935e-01)
        (2.50000e+00, 9.56785e-01)
        (2.75000e+00, 8.63794e-01)
        (3.00000e+00, 6.88224e-01)
        (3.25000e+00, 4.67271e-01)
        (3.50000e+00, 2.59869e-01)
        (3.75000e+00, 1.21738e-01)
        (4.00000e+00, 4.80000e-02)
        (4.25000e+00, 1.70467e-02)
        (4.50000e+00, 6.20561e-03)
        (4.75000e+00, 1.80531e-03)
        (5.00000e+00, 5.34031e-04)
        (5.25000e+00, 1.58431e-04)
        (5.50000e+00, 3.99920e-05)
        (5.75000e+00, 9.45180e-06)
      };
      \addlegendentry{Floating-point}
      \addplot[very thick,color=red,mark=+] plot coordinates {
        (2.00000e+00, 9.99383e-01)
        (2.25000e+00, 9.95402e-01)
        (2.50000e+00, 9.74056e-01)
        (2.75000e+00, 9.04766e-01)
        (3.00000e+00, 7.52972e-01)
        (3.25000e+00, 5.33364e-01)
        (3.50000e+00, 3.14523e-01)
        (3.75000e+00, 1.50056e-01)
        (4.00000e+00, 6.26916e-02)
        (4.25000e+00, 2.25234e-02)
        (4.50000e+00, 7.86916e-03)
        (4.75000e+00, 2.18692e-03)
        (5.00000e+00, 8.09339e-04)
        (5.25000e+00, 2.71003e-04)
        (5.50000e+00, 7.94786e-05)
        (5.75000e+00, 2.30012e-05)
        (6.00000e+00, 6.13610e-06)
      };
      \addlegendentry{Fixed-point: $Q_i.Q_c=5.4$}

    \end{semilogyaxis}
  \end{tikzpicture}\hspace{-5pt}
  \begin{tikzpicture}

    \pgfplotsset{
      grid style = {
        dash pattern = on 0.05mm off 1mm,
        line cap = round,
        black,
        line width = 0.5pt
      },
      label style = {font=\fontsize{9pt}{7.2}\selectfont},
      tick label style = {font=\fontsize{9pt}{7.2}\selectfont}
    }

    \begin{semilogyaxis}[%
      xlabel=$E_b/N_0$ (dB),xtick={2,3,...,6.0},%
      xlabel style={yshift=0.4em},%
      minor x tick num={1},
      xmin=2,xmax=6,%
      ymin=1e-7,ymax=1e-1,%
      ylabel=Bit-error rate, ylabel style={yshift=-1.0em},%
      width=0.53\columnwidth, height=6.55cm, grid=major,%
      mark size=3.0pt,
      mark options=solid]
      
      \addplot[very thick,color=black] plot coordinates {
        (2.00000e+00, 6.96643e-02)
        (2.25000e+00, 6.23495e-02)
        (2.50000e+00, 5.26968e-02)
        (2.75000e+00, 4.05652e-02)
        (3.00000e+00, 2.71511e-02)
        (3.25000e+00, 1.53191e-02)
        (3.50000e+00, 7.14722e-03)
        (3.75000e+00, 2.77382e-03)
        (4.00000e+00, 9.09091e-04)
        (4.25000e+00, 2.66887e-04)
        (4.50000e+00, 7.97995e-05)
        (4.75000e+00, 1.97358e-05)
        (5.00000e+00, 4.95846e-06)
        (5.25000e+00, 1.42964e-06)
        (5.50000e+00, 3.38252e-07)
        (5.75000e+00, 7.79855e-08)
      };
      \addplot[very thick,color=red,mark=+] plot coordinates {
        (2.00000e+00, 7.36113e-02)
        (2.25000e+00, 6.65504e-02)
        (2.50000e+00, 5.72969e-02)
        (2.75000e+00, 4.55158e-02)
        (3.00000e+00, 3.16895e-02)
        (3.25000e+00, 1.85270e-02)
        (3.50000e+00, 9.03671e-03)
        (3.75000e+00, 3.48115e-03)
        (4.00000e+00, 1.20071e-03)
        (4.25000e+00, 3.44859e-04)
        (4.50000e+00, 9.84481e-05)
        (4.75000e+00, 2.32946e-05)
        (5.00000e+00, 7.25374e-06)
        (5.25000e+00, 2.32021e-06)
        (5.50000e+00, 6.46255e-07)
        (5.75000e+00, 1.85492e-07)
        (6.00000e+00, 4.91453e-08)
      };

    \end{semilogyaxis}
  \end{tikzpicture}
  \\
  \ref{unrolled-perf-legend}
	\caption{Impact of LLR quantization on the error-correction performance of the systematic (1024,\,869) polar code decoded by the unrolled decoder implementation.}
	\label{fig:unrolled-perf-cmp}\vspace{-3pt}
\end{figure}

The unrolled decoder is implemented in the low-V$_\text{T}$ technology flavor, and occupies an area of 0.35\,mm$^2$ with a density of 64\%. It is built for a high-rate polar code as, in many applications, the peak throughput is achieved in the best channel conditions with a high-rate code. The underlying assumption is that the unrolled decoder---implementing an \gls{sc}-based algorithm that does not offer as good of an error-correction performance than \gls{scl} or \gls{scf} decoding---would only be used when the channel conditions are good. Thus, the unrolled decoder is built for a systematic (1024,\,869) polar code optimized for $\nicefrac{E_b}{N_0}=4.0$\,dB, and with an initiation interval $\mathcal{I}=50$. It has a fixed latency of 283\,\glspl{cc} and uses $Q_i$.$Q_c=5.4$ to represent \glspl{llr}. Fig.~\ref{fig:unrolled-perf-cmp} shows that using this \gls{llr} quantization leads to a coding loss of under 0.13\,dB at a \gls{fer} of $10^{-4}$ or at a \gls{ber} of $10^{-6}$.
To keep the longest combinational paths balanced, the dedicated decoders for the Repetition and \gls{spc} codes were constrained to a maximum length of 8 and 4, respectively. The critical path starts from the output of an \gls{llr} register, goes through a dedicated decoder for a \gls{spc} code of length 4, and ends at the input of a bit-estimate register.
Instead of using enable signals for the registers, it makes heavy use of clock gating, thus significantly reducing the area and power requirements.

In the following, the measured throughput and energy per bit are presented, and briefly discussed.

\subsubsection{Throughput and Energy-per-bit Comparisons}
The throughput of the unrolled \gls{sc} decoder is over an order of magnitude greater than any of the flexible decoder modes. At a supply voltage of 0.9\,V, its coded throughput is of 9\,234\,Mbps at an achievable clock frequency $f_{\text{clk}}$ of 451\,MHz.
The energy per bit is shown to be of 2.55\,pJ/bit at 100\,MHz or of 1.15\,pJ/bit at 451\,MHz. For this decoder implemented with low-V$_{\text{T}}$ cells, leakage makes for the majority of the total power consumption at 100\,MHz: 3.9\,mW out of 5.2\,mW. At 451\,MHz, the contribution of the leakage drops down to a third of the total power consumption.

\subsubsection{Discussion}
The throughput of the unrolled \gls{sc} decoder is over an order of magnitude than those of the various modes supported by the flexible decoder, as presented in Fig.~\ref{fig:tp-cmp}. Comparing the energy per bit of the two architectures confirms that an unrolled \gls{sc} decoder built for a specific polar code can achieve the lowest energy per bit. This speed and energy-efficiency comes at the expense of flexibility.

\subsection{Comparing with the State-of-the-Art Fabricated ASICs}
Only two other fabricated ASICs can be found in the literature, both are for polar codes with a blocklength $N=1024$. In \cite{Mishra2012}, Mishra et al. presented a rate-flexible \gls{sc} decoder fabricated in UMC's 180\,nm CMOS technology. In \cite{Park2014}, Park et al. presented a rate-flexible \gls{bp} decoder fabricated in TSMC's 65\,nm CMOS technology. The results reported in \cite{Park2014} focus on a (1024,\,512) polar code decoded at a high $\nicefrac{E_b}{N_0}$ value where the average number of iterations is of 6.57 out of the maximum of 15 iterations.

Table~\ref{tab:asic:flexible} shows a comparison of our flexible decoder against the other fabricated ASIC decoders. We present some results for the  three supported modes: \gls{sc}, \gls{scf} with a maximum number of trials $T=8$, and \gls{scl} with a list size $L=4$. An 8-bit \gls{crc} is used for the \gls{scf} and \gls{scl} decoders. We present \gls{scl} results for three different core supply voltages. For fair comparison against \cite{Park2014}, the table focusses on a (1024,\,512) polar code decoded at a $\nicefrac{E_b}{N_0}=4$\,dB. Note that the \gls{fer} at $\nicefrac{E_b}{N_0}=4$\,dB for the \gls{bp} decoder was taken from \cite[Fig.~4.10]{ParkThesis}---the Ph.D. thesis of the first author of \cite{Park2014}. The \glsfirst{wc} coded throughput is also included as some decoding algorithms have a throughput that depends on the channel conditions. Since the results for the state of the art are for other technologies and supply voltages, normalized results are also provided for comparison.

Looking at results for the different modes of the flexible decoder, the same remarks formulated in Sections~\ref{sec:results:tp} and \ref{sec:results:energy} apply when the core voltage is 0.9\,V for all modes. At 0.9\,V, the \gls{sc} decoder shows the lowest latency and greatest throughput. Still at the same core supply, the throughput and energy efficiency of the \gls{scf} mode are on par with the \gls{sc} decoder when the $\nicefrac{E_b}{N_0}$ ratio is sufficiently high, i.e., when the number of trials becomes approximately 1. The \gls{scl} mode trails behind but still remains within the same order of magnitude.

Comparing our flexible decoder with the normalized results for the other works, it can be seen from Table~\ref{tab:asic:flexible} that the \gls{bp} decoder of \cite{Park2014} has the lowest latency and greatest throughput while the \gls{sc} decoder of \cite{Mishra2012} has the smallest area and best energy efficiency. It should be noted however that the error-correction performance of the \gls{bp} decoding algorithm is significantly worse than that of any of the three algorithms supported by our flexible decoder, and that the decoder of \cite{Mishra2012} is specialized for \gls{sc} decoding. Our flexible decoder is not optimized for efficient \gls{sc} decoding, it implements the \gls{sc} algorithm by using parts of the \gls{scl} decoder. Similarly, the area efficiency results for the \gls{sc} and \gls{scf} modes are not suitable for a fair comparison against the other works as these two modes use only a fraction of the flexible decoder area, an area dictated by the largest list size supported by the \gls{scl} mode.

\begin{table}[t]
  \centering
  \setlength{\tabcolsep}{5pt}
  \caption{Comparison of the unrolled decoder against the other fabricated ASIC decoders for a (1024,\,869) polar code.}
    \begin{tabular}{l|c|c|cc}
      \toprule\hline
      \textbf{Implementation}& \textbf{This work} & \cite{Mishra2012} & \multicolumn{2}{c}{\cite{Park2014}}\\
      \hline\hline
      \textbf{Algorithm}     & SC & SC   & \multicolumn{2}{c}{BP (15 iter.)}\\\hline
      \textbf{$\nicefrac{E_b}{N_0}$ @ FER of $10^{-3}$}&4.95&4.95&\multicolumn{2}{c}{5.20}\\\hline
      \textbf{Technology}          & 28\,nm     & 180\,nm &\multicolumn{2}{c}{65\,nm}  \\\hline
      \textbf{Area} (mm$^2$)       & 0.35      & 1.71  & \multicolumn{2}{c}{1.48}\\\hline
      \textbf{Supply} (V)          & 0.9       & 1.3   & 1.0 & 0.475 \\\hline
      \textbf{Frequency} (MHz)     &451& 150   & 300 & 50 \\\hline
      \multirow{2}{*}{\textbf{Latency} }\multirow{2}{*}{\Shortunderstack{(CCs)\\($\mu$s)}}%
                                   & 283       & 1\,568  & \multicolumn{2}{c}{150}\\\cline{2-5}
                                   &0.63& 10.45 & 0.50 & 3.00 \\\hline
      \textbf{W.-C. Coded T/P} (Mbps)&9\,233.8& 98.0   & 2\,048.0 & 341.3 \\\hline
      \textbf{Area Eff.} (Mbps/mm$^2$)&26\,741& 57    & 1\,384 & 231\\\hline
      \textbf{Power} (mW)          &10.6& 67    & 477.5 & 18.6\\\hline
      \textbf{Energy per bit} (pJ/bit)&1.2& 684   & 233  & 54\\\hline
      \hline
      \multicolumn{5}{l}{\textit{Normalized for 28\,nm and 0.9\,V$\,^\diamond$}}\\
      \hline\hline
      \textbf{Area} (mm$^2$)          & 0.35&0.04&\multicolumn{2}{c}{0.27}\\\hline
      \textbf{Frequency} (MHz)        &451&1\,335&696 &--\\\hline
      \textbf{Latency} ($\mu$s)       &0.63&1.17&0.22&--\\\hline
      \textbf{W.-C. Coded T/P} (Mbps) &9\,233.8&871.9& 4\,751.4 & -- \\\hline
      \textbf{Area Eff.} (Mbps/mm$^2$)&26\,741&21\,073&17\,301&--\\\hline
      \textbf{Power} (mW)             &10.6&5.0 & 166.6 & -- \\\hline
      \textbf{Energy per bit} (pJ/bit)&1.2&5.7&35.1&--\\\hline
      \bottomrule
      \multicolumn{5}{l}{$^\diamond$Area scaled as $s^2$, frequency as $\nicefrac{1}{s}$, and power as $v^2s$, where $s$ is the}\\
      \multicolumn{5}{l}{\phantom{$^\diamond$}technology feature size and $v$ is the supply voltage ratio.}\\
      \multicolumn{5}{l}{\phantom{$^\diamond$}The frequency of \cite{Mishra2012} was first scaled back linearly to $1.8$\,V, the nom-}\\
      \multicolumn{5}{l}{\phantom{$^\diamond$}inal voltage of the 180\,nm technology.}\\
    \end{tabular}
  \label{tab:asic:unrolled}
\end{table}

Table~\ref{tab:asic:unrolled} compares the measurement results for our dedicated unrolled decoder, specialized for one polar code, against those of the same two fabricated rate-flexible decoders \cite{Mishra2012,Park2014}. Note that by lack of data, and for fair comparison, we present worst-case throughput results for the \gls{bp} decoder.
Similarly to Table~\ref{tab:asic:flexible}, normalized results are presented. 
Comparing solely with the normalized results, it can be seen that the unrolled decoder outperforms the other works in terms of throughput and energy efficiency for an area efficiency in the same vicinity. Compared to the normalized results of the other \gls{sc} decoder, the area of our decoder is approximately 10$\times$ greater, however the throughput is also 10$\times$ greater and the latency 1.8$\times$ lower. The area of our decoder is $1.3\times$ that of the normalized area for the \gls{bp} decoder, the throughput near double and the latency approximately three times greater. The energy per bit of our decoder was measured to be 4.75$\times$ and 29.25$\times$ smaller than the normalized energy-per-bit values of \cite{Mishra2012} and~\cite{Park2014}, respectively.

\subsection*{Further Discussion}
\vspace{3pt}

We note that the field of polar codes has been very active since the RTL of \polarbear has been finalized. Many improvements were proposed to the \gls{scl} decoding algorithm and its implementation in particular. Notably, more efficient \glspl{psn} were proposed in \cite{Fan2014}, multi-bit and tree pruning methods presented~\cite{Sarkis_JSAC_2016,Yuan2015}, or a combination of both, e.g.~\cite{Lin2015c,Fan2016}. These improvements are orthogonal to our work.

\begin{table}[t]
  \centering
  \setlength{\tabcolsep}{5pt}
  \caption{Synthesis-result comparison of SCL decoders for a (1024,\,512) polar code.}
  {
    \begin{tabular}{l|c|c|c}
      \toprule\hline
      \textbf{Implementation}& \textbf{This work} & \cite{Lin2015c} & \cite{Fan2016}\\
      \hline\hline
      \textbf{List size}          & 4     & 4 & 16  \\\hline
      \textbf{Technology}          & 28\,nm     & 90\,nm & 90\,nm  \\\hline
      \textbf{Area} (mm$^2$)       & 0.3       & 3.83  & 7.47\\\hline
      \textbf{Frequency} (MHz)     & 500 & 403 & 658 \\\hline
      \multirow{2}{*}{\textbf{Latency} }\multirow{2}{*}{\Shortunderstack{(CCs)\\($\mu$s)}}%
                                   & 2\,408   & 371  & 1\,462\\\cline{2-4}
                                   & 4.82 & 0.92 & 2.22 \\\hline
      \textbf{Coded T/P} (Mbps)& 212.6 & 1\,112.3 & 460.9 \\\hline
      \textbf{Area Eff.} (Mbps/mm$^2$)& 709 & 290 & 62 \\\hline
      \hline
      \multicolumn{4}{l}{\textit{Normalized for 28\,nm and list size $L=4$$\,^\diamond$}}\\
      \hline\hline
      \textbf{Area} (mm$^2$)       & 0.3    & 0.4    & 0.2 \\\hline
      \textbf{Frequency} (MHz)     & 500    & 1\,295 & 2\,115 \\\hline
      \textbf{Latency} ($\mu$s)    & 4.82   & 0.29   & 0.69 \\\hline
      \textbf{Coded T/P} (Mbps)    & 213    & 3\,575 & 1\,481 \\\hline
      \textbf{Area Eff.} (Mbps/mm$^2$)& 709 & 9\,645 & 8\,195 \\\hline
      \bottomrule
      \multicolumn{4}{l}{$^\diamond$Area scaled as $s^2l$, and frequency as $\nicefrac{1}{s}$, where $s$ is the}\\
      \multicolumn{4}{l}{\phantom{$^\diamond$}technology feature size and $l$ is the list-size ratio.}\\
    \end{tabular}
    }
  \label{tab:synth:scl}
\end{table}

\vspace{7pt}

To help estimate the potential impact that could be brought by recent architectural improvements, Table~\ref{tab:synth:scl} presents a comparison between our synthesis results for our flexible decoder (with emphasis on the \gls{scl} mode) against those from the state of the art works of \cite{Lin2015c,Fan2016}. Normalized results, including to account for the different list size of \cite{Fan2016}, are presented.

\vspace{7pt}

Comparing the latency in \glspl{cc} of our decoder with the other works, it can be seen that the reduced-latency algorithm of \cite{Lin2015c}, that notably estimates multiple bits at once, can have a significant impact. The approximate metric sorter of \cite{Fan2016} also leads to a latency reduction. Looking at the normalized results, it can be seen that the area results are in the same vicinity. The improved \gls{psn} of \cite{Lin2015c,Fan2016} and the approximate sorter of \cite{Fan2016} lead to much greater clock frequencies. By comparing the achievable clock of our synthesized design with that of our on-chip flexible decoder at 0.9\,V (Table~\ref{tab:asic:flexible}) hints that the gains that are expected from standard scaling laws appear to be difficult to fully realize, especially with regular-V$_{\text{T}}$ libraries. This is partly due to the impact of parasitics and wiring.

\vspace{7pt}

A detailed survey that includes the recent work and a comparison of polar decoders with \gls{ldpc} and Turbo decoders can be found in~\cite{Balatsoukas_WCNC_2017}. The comparison discusses, among other things, the required list size  and blocklength for \gls{scl} decoding in order to match the performance of various \gls{ldpc} and Turbo decoders. Another important implementation-related aspect is the quantization loss, which we showed in Section~\ref{sec:results} to be negligible when using bit-widths that are very similar to the bit-widths commonly used in \gls{ldpc} decoders.

\vfill

\section{Conclusion}\label{sec:conclusion}
In this paper, we presented measurement results for \polarbear, a fabricated chip in 28\,nm FD-SOI CMOS technology that implements two decoders for polar codes. The first decoder is flexible, supporting three different modes corresponding to distinct decoding algorithms: \gls{sc}, \gls{scf} and \gls{scl}. It implements a latency saving technique applicable to all three decoding algorithms. Furthermore, this flexible decoder can decode both non-systematic and systematic polar codes of any code rate. The list size and maximum number of trials for \gls{scl} and \gls{scf} decoding modes, respectively, are configured at execution time. The second decoder is a fully-unrolled partially-pipelined \gls{sc} decoder built for speed. To our knowledge, this paper presents the first ASIC measurement results for both the \gls{scf} and \gls{scl} algorithms.

We presented a flexible decoder where the most suitable mode can be selected at execution based on the requirements and operating conditions. 
The \gls{sc} mode was shown to have the best throughput and energy per bit.
For the best error-correction performance, the \gls{scl} mode was shown to be the most favorable choice at the expense of a greater energy per bit compared to both the \gls{sc} and \gls{scf} modes. Lastly, with an error-correction performance that approaches that of the \gls{scl} algorithm with $L=2$ and an average throughput that tends to that of the \gls{sc} mode as the signal-to-noise ratio improves, the \gls{scf} mode appeared as the most attractive mode if the decoder is operated in a good $\nicefrac{E_b}{N_0}$ region and if the system can cope with a variable execution time.

In terms of more specific results, we showed that in the \gls{scl} mode, our flexible decoder could achieve a coded throughput of 306.8\,Mbps with a latency of 3.34\,$\mu$s and an energy per bit of 418.3\,pJ/bit at a clock frequency of 721\,MHz for a supply of 1.3\,V. The energy efficiency was shown to improve twofold with the energy per bit dropping down to 178.1\,pJ/bit with the more modest clock frequency of 308\,MHz, throughput of 130.9\,Mbps, and supply voltage of 0.9\,V. In the other two operating modes, our measurement results showed that our flexible decoder had an energy per bit of approximately 95\,pJ/bit with a core supply voltage of 0.9\,V. It should be noted that research on \gls{scl} decoding has been moving quickly since this chip has been sent out for tapeout. This makes us confident that ASIC results can only improve from here if all the latest improvements from the recent literature are to be incorporated.

The unrolled decoder was shown to be capable of achieving an area efficiency of 26.74\,Gbps/mm$^2$ at 451\,MHz for a supply voltage of 0.9\,V and an energy per bit of 1.15\,pJ/bit. These results confirmed that a specialized unrolled polar decoder has significantly better energy per bit and speed than its flexible counterpart. When it comes to the energy-efficiency advantage, the key ingredients were the polar code-specific specialization, the unrolling, and the use of clock gating. Since the design of this chip, it was shown in \cite{Giard_TCAS_2016} that such a decoder can be made to support multiple codes of various rates and blocklengths.

\vspace{-4pt}
\section*{ACKNOWLEDGEMENT}
The authors would like to thank Christian Senning and Lorenz Schmid (formerly EPFL) for their support, Ivan Miro-Padanes (CEA-LETI) for providing the \gls{fll}~\cite{Miro2014} along with support for it, and Marc-Andr\'e Carbonneau (\'ETS) for the test PCB design. Furthermore, they would like to thank STMicroelectronics for chip fabrication.

\vspace{-4pt}


\end{document}